\documentclass[sigconf,nonacm]{acmart}

\AtBeginDocument{%
  }
\usepackage{algorithm}
\usepackage[noend]{algpseudocode}

\usepackage{xcolor}
\usepackage{enumitem}
\usepackage{afterpage}
\makeatletter
\newcommand\myparagraph{%
  \def\@toclevel{4}%
  \@startsection{paragraph}{4}{0pt}%
    {-.4\baselineskip \@plus -2\p@ \@minus -.2\p@}%
    {-3.5\p@}%
    {\ACM@NRadjust{\normalfont\bfseries\@adddotafter}}%
}
\makeatother

\begin{document}

\title{The Blind Spots of Detection: Adversarial Perturbation Attacks against Watchdogs in Distributed LLM Inference}

\author{Siu Wai Shi}
\affiliation{%
  \institution{Shanghai Jiao Tong University}
  \city{Shanghai}
  \country{China}
}

\author{Xiaochen Zhu}
\affiliation{%
  \institution{Massachusetts Institute of Technology}
  \city{Cambridge}
  \state{Massachusetts}
  \country{USA}
}

\author{Xinjian Luo}
\affiliation{%
  \institution{Shanghai Jiao Tong University}
  \city{Shanghai}
  \country{China}
}

\author{Keke Huang}
\affiliation{%
  \institution{Huazhong University of Science and Technology}
  \city{Wuhan}
  \state{Hubei}
  \country{China}
}

\author{Bingxue Zhang}
\affiliation{%
  \institution{University of Shanghai for Science and Technology}
  \city{Shanghai}
  \country{China}
}

\author{Xiaofeng Gao}
\affiliation{%
  \institution{Shanghai Jiao Tong University}
  \city{Shanghai}
  \country{China}
}

\renewcommand{\shortauthors}{Shi et al.}

\begin{abstract}
Decentralizing large language model (LLM) inference across untrusted Web nodes requires robust verification to ensure computational integrity.
Existing activation-based verification and detection mechanisms validate intermediate outputs against reference values, permitting bounded numerical deviations to accommodate benign cross-hardware discrepancies.
However, a critical security question remains: what can a malicious worker achieve while operating within these tolerance bounds?

In this work, we introduce \textsc{Hush-Up}, a stealthy availability attack against distributed LLM inference, where a single malicious worker halts generation while evading attribution.
The attacker first profiles its local execution to model an empirical bound of benign hardware errors. Using this error bound as a feasibility constraint, the attacker optimizes an adversarial perturbation on its output activations to maximize the final-layer probability of the end-of-sequence (EOS) token, i.e., to \emph{hush up} the model.
Across diverse models and benchmarks, \textsc{Hush-Up} effectively forces premature generation termination and collapses model capabilities.
Furthermore, the attack \emph{hushes up} the watchdogs themselves:
existing defenses often fail to detect the sabotage entirely or falsely attribute the amplified downstream anomalies to honest workers.
Our findings expose a fundamental gap between local numerical verification and end-to-end accountability: perturbations too small to indict a worker locally can nevertheless be powerful enough to break distributed inference globally.

\end{abstract}

\settopmatter{printacmref=false}


\maketitle

\section{Introduction}

\begin{figure*}[t]
    \centering
    \includegraphics[width=0.90\textwidth]
        {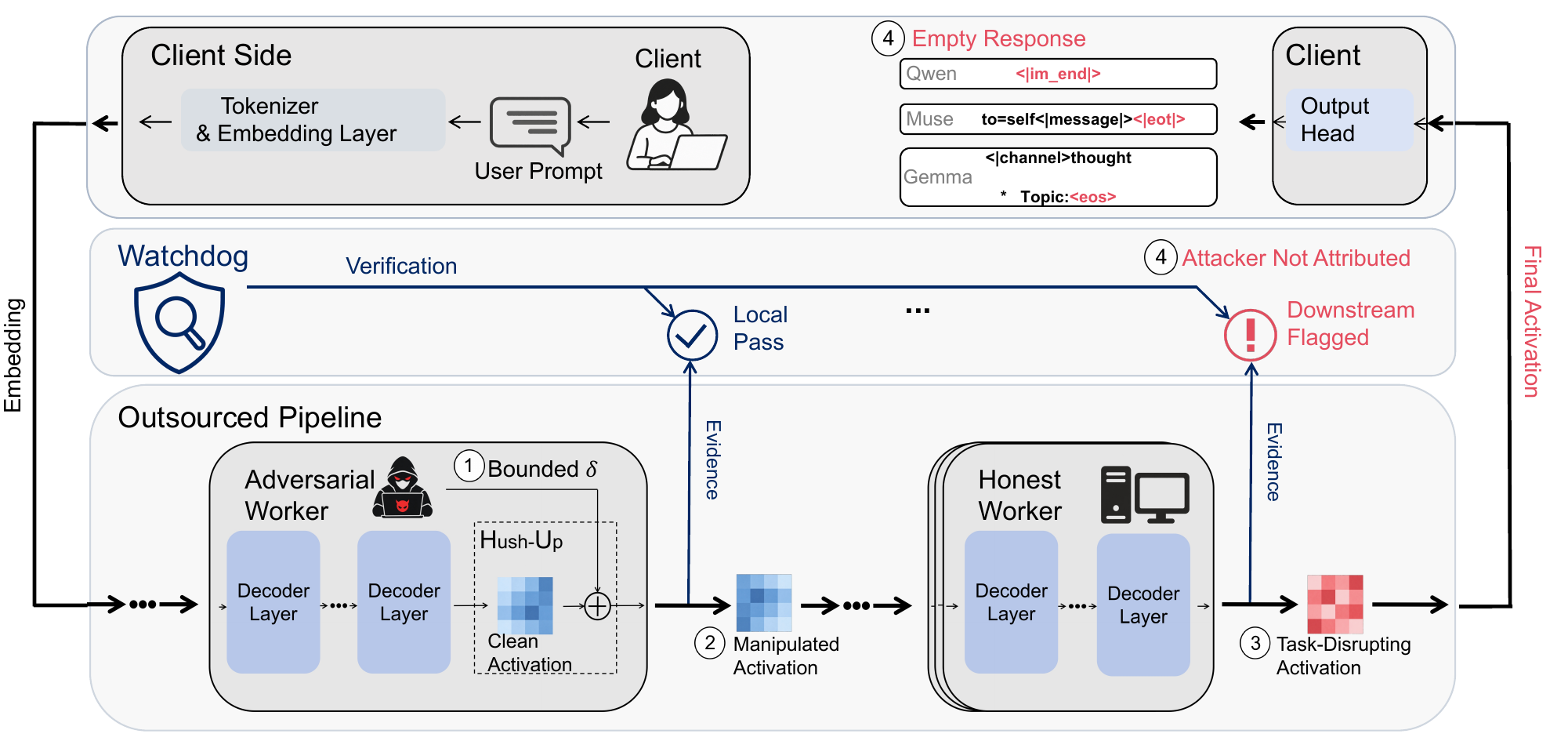}
    \vspace{-\baselineskip}
    \caption{Overview of \textsc{Hush-Up}, illustrating
    an attribution-failure scenario. A malicious worker
    injects a bounded perturbation into its prefill
    activations. The perturbation passes the local check
    but disrupts downstream computation and prematurely
    terminates generation, while the malicious worker
    escapes correct attribution.}
    \label{fig:attack-overview}
    \vspace{-\baselineskip}
\end{figure*}

Many of today's large language models (LLMs) have outgrown commodity hardware: their memory footprints often exceed the capacity of any consumer-grade GPU, and even when they can be made to fit (e.g., by offloading weights to CPU memory or even disk~\cite{pmlr-v202-sheng23a}), inference becomes prohibitively slow~\cite{song2024powerinferfastlargelanguage}.
This has pushed LLM inference beyond individual machines and onto the open Web. Rather than relying on a single provider's data center, distributed inference~\cite{borzunov2023petals, borzunov2023distributed, tong2025parallaxefficientllminference, chen2026distributedgenerativeinferencellm, pmlr-v235-jiang24f,wu2025deserveaffordableofflinellm} democratizes access to state-of-the-art open-weight LLMs by allowing independent participants to pool their own machines. It splits the model into consecutive stages hosted by different parties, so that together they can serve a model that none of them could run alone.
However, the same openness removes the usual grounds for trust.
When different parts of the model are hosted by unknown operators across the Web, no participant can simply be assumed to behave honestly.
Distributed inference on the Web therefore depends on \emph{verification}: checking that each participant executed its share of the computation faithfully and, when one did not, identifying who is responsible.

In principle, cryptographic proofs of correct execution provide strong integrity guarantees~\cite{10.1145/3627703.3650088, kang2022scalingtrustlessdnninference, NIPS2017_6048ff4e}, but their substantial overhead is prohibitive at the scale of modern LLMs~\cite{sun2024zkllm,qu2025zkgpt, ong2025toploc}.
Practical verification and detection mechanisms, which we call \emph{watchdogs}, therefore take a lighter-weight approach.
They validate the intermediate activations, 
or compact fingerprints of them, against reference values that are typically obtained by re-executing the prescribed computation or established in advance~\cite{ong2025toploc, karvonen2025difr,yao2026tao,dolatabadi2026sentinel, cihangiroglu2026integritypeertopeerdistributedllm}.
These checks cannot require exact matches: because of floating-point rounding, honest executions of the same model on the same input produce slightly different activations across heterogeneous hardware~\cite{yuan2025understanding, zhang2025hardwaresoftwareplatforminference, wimbauer2026fingerprintinginferencesystemslarge}.
Therefore, every activation-based watchdog must tolerate small numerical deviations and accept them as benign.

This necessary tolerance, however, leaves room for a malicious worker and raises a crucial question: \emph{if a deviation is too small to blame, is it also too small to matter?}
There are two reasons to suspect not.
First, deep networks can amplify small perturbations:
\citet{zamir2026noncomposability} shows that locally tolerated deviations can, in theory, steer a network's final output toward an arbitrary target. Second, even when such a deviation is eventually caught, the blame may land on the wrong worker. As honest downstream workers faithfully propagate the perturbed activations, the discrepancy can compound across stages until it triggers an alert later in the pipeline. The watchdog may then blame an honest downstream worker, while the malicious worker that actually injected the perturbation goes unidentified.
Yet the extent to which such bounded perturbations can degrade an LLM's real-world capabilities, and evade detection or attribution by existing watchdogs, remains largely unexplored.

In this paper, we fill this gap and show that both happen in practice.
Specifically, we consider distributed LLM inference with a single malicious worker.
This worker manipulates the intermediate activations it passes downstream, while all other workers execute the prescribed computation honestly.
The attacker aims to tamper with the inference and degrade the model's capabilities without being identified.
Blindly shifting the predicted token probabilities, however, does not necessarily make the model any less capable.
We therefore focus on the most unambiguous form of degradation: terminating generation before any substantive response is produced, which fails any task under any metric.
On the open Web, this amounts to an unaccountable \emph{denial of service} where clients receive no useful responses.
Because the culprit is not identified, it cannot be excluded and remains free to sabotage further requests.

To induce this failure, we develop \textsc{Hush-Up}, an adversarial perturbation attack designed to \emph{hush up} the model.
The malicious worker first measures how much honest executions of its stage differ across GPUs and adopts this benign discrepancy as its perturbation budget.
Within this budget, it then optimizes a one-time perturbation of the prefill activations and sends the perturbed ones to the downstream worker.
The perturbation maximizes the probability that the model emits an end-of-sequence (EOS) token, either as the first generated token or immediately after the model-specific header that opens its response.
Figure~\ref{fig:attack-overview} illustrates the attack
and a representative attribution-failure scenario.

Our evaluation reveals that these optimized perturbations reliably silence recent open-weight LLMs. Across three model families (Qwen, Muse, and Gemma) and three task domains (math, code, and instruction following), \textsc{Hush-Up} reduces task success from 70.83--98.33\% under honest execution to an average of 2.96\%, with 92.31\% of attacked generations terminating within the first 30 tokens.
By contrast, random perturbations of the same magnitude leave generation and task performance essentially intact. This indicates that a deviation of benign scale is harmless unless it is adversarially chosen.
Moreover, \textsc{Hush-Up} is robust across deployment settings.
It remains effective regardless of where the malicious worker sits in the pipeline.
It is likewise effective across three hardware configurations, and perturbations optimized on one configuration transfer to the others without re-optimization.

Existing watchdogs, in turn, fail to hold the malicious worker accountable.
We evaluate four recent activation-based watchdogs spanning fingerprint-based, statistical, and operator-level verification approaches.
In 10 of the 12 model--watchdog configurations in our main experiments, the watchdog \emph{never} correctly identifies the malicious worker.
This failure takes two forms.
In some cases, the watchdog accepts the corrupted execution outright; in others, it detects that something went wrong but is unable to correctly locate its source.
In the latter case, the deviation becomes visible only downstream of where it was injected, so the blame falls on honest workers or on no one.
\textsc{Hush-Up} thus hushes up not only the language model, but also the watchdogs meant to guard it.

These results expose a gap between local numerical verification and end-to-end accountability: a deviation too small to blame where it is injected can still be large enough to break the inference as a whole.
Left unaddressed, this gap undermines the promise of distributed inference on the Web: anyone can contribute compute, yet a single participant can sabotage the service without consequence.
Watchdogs must therefore reason about how deviations propagate across workers, rather than judge each stage in isolation.
We discuss the implications for watchdog design in Section~\ref{sec:discussions}.

Our main contributions can be summarized as follows:
\begin{enumerate}[nosep,wide]
\item \textbf{Exposing a security gap in distributed LLM inference.}
We show that the numerical tolerance that activation-based watchdogs must grant to honest execution is itself an attack surface:
perturbations within this tolerance can still break LLM inference without being attributed. This reveals that local activation-based verification is insufficient to secure distributed inference.

\item \textbf{Designing \textsc{Hush-Up}.}
We develop a watchdog-agnostic attack in which a single malicious worker, using a budget calibrated from benign cross-hardware variation, perturbs its output activations to force premature termination, without modifying model weights or controlling any other worker.

\item \textbf{Evaluating task degradation and accountability.} 
We systematically evaluate \textsc{Hush-Up} across multiple model families, tasks, worker positions, and hardware configurations, and demonstrate substantial task degradation. We further evaluate \textsc{Hush-Up} against existing activation-based watchdogs, characterizing their ability to detect and attribute malicious deviations in distributed inference.

\end{enumerate}

\section{Preliminaries}
\label{sec:prelim}

\subsection{Distributed LLM Inference}
\label{sec:distributed-inference}

An LLM maps a sequence of input tokens $x=(x_1,\ldots,x_n)$ over a vocabulary $\mathcal{V}$ to a probability distribution $p(\cdot\mid x)$ over the next token.
It consists of an embedding layer, a stack of transformer blocks, and an output head, i.e., a final normalization layer and a linear projection followed by a softmax over $\mathcal{V}$~\cite{vaswani2023attentionneed}.

\begin{figure}[t]
    \centering
    \includegraphics[trim={55pt 0pt 25pt 0pt}, clip,width=\columnwidth]
        {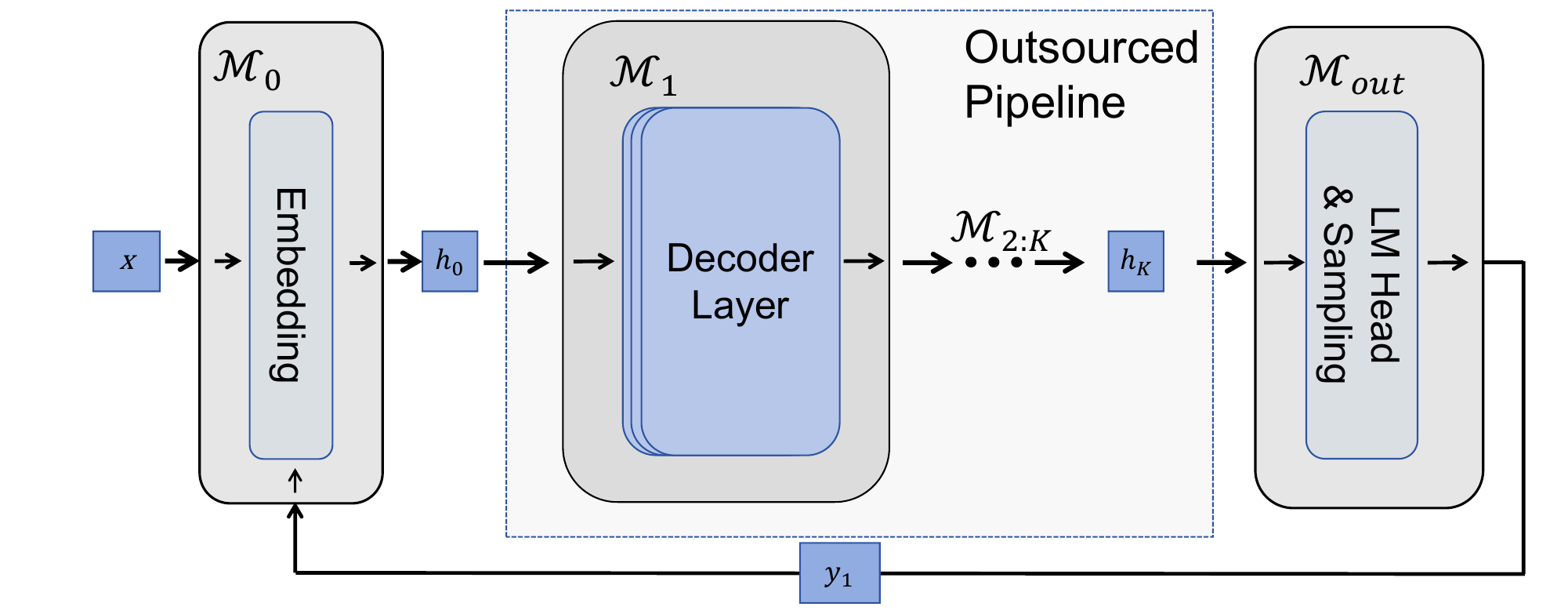}
    \vspace{-\baselineskip}
    \caption{Distributed LLM inference. The client retains
    the embedding layer and output head, while workers
    execute consecutive stages of transformer blocks.
    Generated tokens are fed back into the pipeline
    for autoregressive decoding.}
    \label{fig:distributed-inference}
    \vspace{-\baselineskip}
\end{figure}

In distributed inference~\cite{borzunov2023petals,borzunov2023distributed}, the transformer blocks are partitioned into $K$ consecutive stages $\mathcal{M}_1,\ldots,\mathcal{M}_K$, and each stage $\mathcal{M}_i$ is hosted by a different worker $W_i$.
The client that issues the request retains the embedding layer $\mathcal{M}_0$ and the output head $\mathcal{M}_{\mathrm{out}}$, so the model can be written as
$p(\cdot\mid x)=\mathcal{M}_{\mathrm{out}}\circ\mathcal{M}_K\circ\cdots\circ\mathcal{M}_1\circ\mathcal{M}_0(x)$.
To serve a request, the client computes $h_0=\mathcal{M}_0(x)$ and sends it to $W_1$.
Each worker $W_i$ then computes its output activation
    $h_i=\mathcal{M}_i(h_{i-1})$
and sends it to $W_{i+1}$, or back to the client if $i=K$.
Each activation $h_i\in\mathbb{R}^{n\times d}$ holds one $d$-dimensional hidden state per input position, and the client obtains $p(\cdot\mid x)=\mathcal{M}_{\mathrm{out}}(h_K)$.
Figure~\ref{fig:distributed-inference} illustrates this
execution pipeline and the autoregressive feedback loop.

The response is generated autoregressively in two phases.
In the \emph{prefill} phase, a single forward pass processes the entire prompt, and the client samples the first token $y_1\sim p(\cdot\mid x)$.
In the subsequent \emph{decoding} phase, the client feeds each generated token back into the pipeline and samples $y_t\sim p(\cdot\mid x,y_{<t})$, until the model emits a designated EOS token or a length limit is reached.
Each decoding pass processes only the newly generated token, so the workers exchange a single $d$-dimensional hidden state per step; to attend to earlier positions, each worker reuses the keys and values cached in its attention layers (i.e., the KV cache).
As a result, the activations exchanged during prefill influence every generated token.

\subsection{Numerical Variation across Hardware}
\label{sec:numerical-variation}

Honest executions of the same model on the same input do not necessarily produce identical activations~\cite{yuan2025understanding, gond2026llm42enablingdeterminismllm}.
Floating-point arithmetic has finite precision, and floating-point addition is not associative: for some values of $a$, $b$, and $c$, $\operatorname{fl}(a+\operatorname{fl}(b+c))\ne\operatorname{fl}(\operatorname{fl}(a+b)+c)$, where $\operatorname{fl}(\cdot)$ denotes floating-point evaluation.
GPU architectures and kernel implementations differ in how they order reductions, tile computations, and fuse operations in matrix multiplications and normalizations, so the same computation is rounded differently on different hardware~\cite{yuan2025understanding}.
These rounding errors are larger under lower precision~\cite{yuan2025understanding}, such as BF16, in which LLM activations are typically represented~\cite{nvidia2023tensorrtllm}. Figure~\ref{fig:hardware-pca} visualizes intermediate activations produced by three GPUs at matched token positions.

\afterpage{%
  \begin{figure}[t]
    \centering
    \includegraphics[width=\columnwidth]{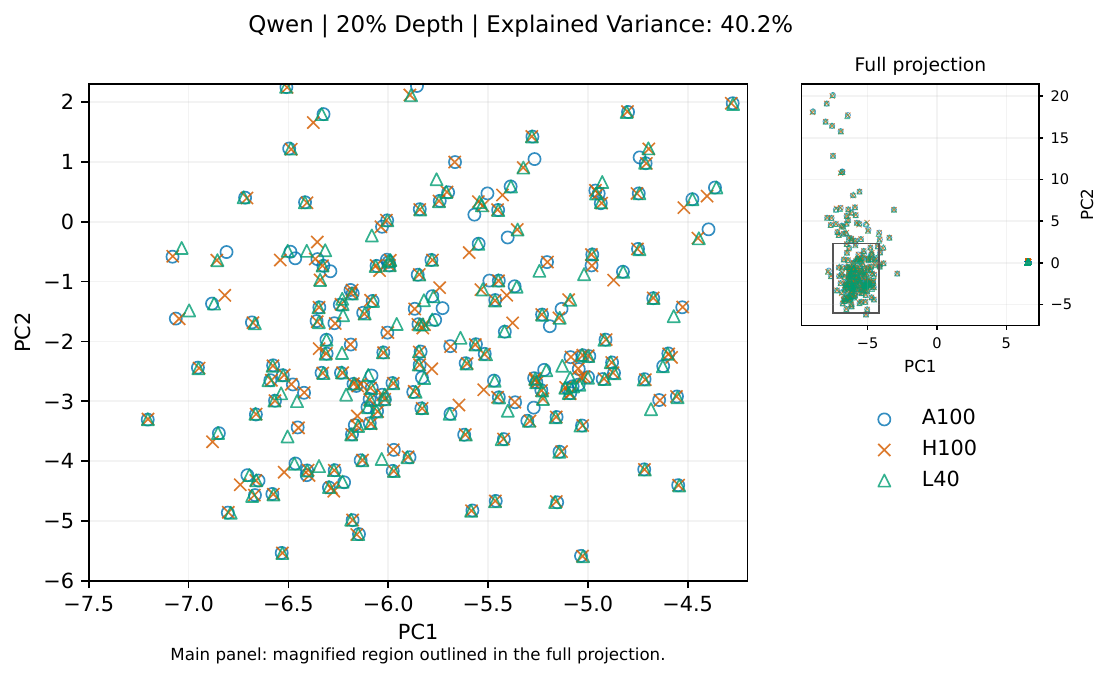}
    \vspace{-\baselineskip}
    \caption{
        PCA projection of intermediate activations at
        425 matched token positions.
        Corresponding activations cluster closely
        but exhibit small numerical differences.
        }
    \label{fig:hardware-pca}
    \vspace{-\baselineskip}
\end{figure}

}

To quantify this variation at the output of stage $\mathcal{M}_i$, consider two executions of the model on the same input under environments $a$ and $b$ (e.g., two GPU types) with identical weights, numerical precision, and inference configuration, producing activations $h_i^{(a)}$ and $h_i^{(b)}$.
We measure their discrepancy by the relative distance
\begin{equation}
    d(h_i^{(a)},h_i^{(b)})=\|h_i^{(b)}-h_i^{(a)}\|_F/\|h_i^{(a)}\|_F,
    \label{eq:relative-distance}
\end{equation}
where $\|\cdot\|_F$ denotes the Frobenius norm, i.e., the $\ell_2$ norm of the flattened activation tensor.

\subsection{Activation-Based Watchdogs}
\label{sec:watchdogs}

A watchdog verifies distributed inference by examining the intermediate activations.
We model it as a function
\begin{equation}
    \mathcal{W}(e_1,\ldots,e_K)\in\{\textsf{accept}\}\cup\{1,\ldots,K\}\cup\{\bot\},
    \label{eq:watchdog}
\end{equation}
where $e_j$ is the evidence submitted by worker $W_j$.
This evidence is derived from activations associated with $W_j$ over the course of the inference, during both prefill and decoding.
Depending on the specific watchdog design, it may take many forms, such as the activations themselves, selected features of them, compact fingerprints or commitments.
$\mathcal{W}$ compares the evidence against reference information that it obtains independently of the workers, e.g., by re-executing stages or from statistics established in advance.
It then either accepts the execution, rejects it and blames a worker $j\in\{1,\ldots,K\}$, or rejects it without blaming any worker ($\bot$).
We describe the four watchdogs evaluated in this paper in Section~\ref{sec:eval-setup}.

Ideally, a watchdog should satisfy two properties.
\emph{Completeness}: if all workers execute their stages faithfully, $\mathcal{W}$ outputs \textsf{accept} with high probability.
\emph{Soundness}: if a worker $W_i$ deviates from its prescribed computation, $\mathcal{W}$ rejects the execution and correctly attributes the deviation to $W_i$, i.e., outputs $i$.
Both properties are essential, but they can be in tension. When a malicious worker perturbs its activation within the range of honest hardware variation (cf. Section~\ref{sec:numerical-variation}), completeness favors acceptance, whereas soundness requires it to reject. Existing watchdogs usually resolve this tension in favor of completeness, treating deviations within honest variation as benign~\cite{ong2025toploc,yao2026tao,karvonen2025difr}.

\section{Problem Statement}\label{sec:problem}

\myparagraph{System Model}
We consider the distributed inference setting of Section~\ref{sec:prelim}, with $K$ workers that may use heterogeneous hardware and a watchdog $\mathcal{W}$.
To distinguish what a worker should compute from what it actually sends, let $\hat{h}_j$ denote the prefill activation that $W_j$ transmits (with $\hat{h}_0=h_0$), and let
$h_j^f=\mathcal{M}_j(\hat{h}_{j-1})$
denote the \emph{faithful} activation that $W_j$ would obtain on its received input.
An honest worker transmits $\hat{h}_j=h_j^f$, up to benign numerical variation.
Given a malicious worker $W_i$, the output of $\mathcal{W}$ falls into one of three outcomes: a \emph{detection failure} if it is \textsf{accept}, an \emph{attribution success} if it is $i$, and an \emph{attribution failure} otherwise, i.e., if $\mathcal{W}$ rejects the execution but blames an honest worker or no worker at all.

\myparagraph{Threat Model}
We assume that the adversary controls exactly one worker $W_i$ ($1\le i\le K$), while all other parties operate as prescribed.
That is, the client, the other workers, and the watchdog all follow the protocol faithfully.
We further assume an authenticated channel between every pair of parties, so that no party can tamper with the activations or evidence transmitted by another.

The adversary may manipulate the activations that $W_i$ sends to the next worker, without modifying the underlying model weights.
Other than this, $W_i$ follows the protocol. In particular, it submits evidence computed from the activations it actually transmits, so that the watchdog observes the same activations as the next worker.

The adversary has white-box access to the model, including the architecture and weights of all stages and the output head.
This is realistic in practice, since distributed inference primarily serves open-weight LLMs.
The adversary also knows the partition configuration, i.e., how the model is split into stages and which stage it hosts.
However, it does not know the watchdog's algorithm, reference information, or thresholds.
Finally, the adversary has sufficient computational resources and can run its own stage on different types of hardware. The adversary observes only the incoming activations
and does not have access to the original input tokens.

The adversary aims to degrade the model's capabilities.
Specifically, it seeks to silence the model by inducing an EOS token immediately or shortly after generation begins, so that the model fails to complete the requested task.
Meanwhile, it seeks to avoid being identified for its malicious behavior, i.e., such that $\mathcal{W}(e_1,\ldots,e_K)\neq i$, which results in either a detection failure or an attribution failure.

\section{The \textsc{Hush-Up} Attack}

\subsection{Attack Overview}

\textsc{Hush-Up} casts the adversary's goal as a constrained optimization problem.
After computing its faithful prefill activation $h_i^f$, the adversary transmits
\begin{equation}
    \hat{h}_i=h_i^f+\delta_i,
    \label{eq:transmitted-activation}
\end{equation}
where the perturbation $\delta_i$ maximizes an EOS-promoting objective $J$ subject to a budget $\rho_i$ on the relative deviation from the faithful activation:
\begin{equation}
    \max_{\delta}\; J(\delta)
    \quad\text{s.t.}\quad
    d\big(h_i^f,\,\hat{h}_i\big)\le\rho_i.
    \label{eq:attack-budget-constraint}
\end{equation}
The budget limits the perturbation magnitude to the numerical variation observed across faithful executions on different hardware, while the objective steers the perturbation toward immediate or near-immediate termination of generation.
Section~\ref{sec:constraints-objectives} describes how the adversary calibrates $\rho_i$ and defines $J$, and Section~\ref{sec:optimization} describes how it solves Equation~\eqref{eq:attack-budget-constraint}.
Algorithm~\ref{alg:hush-up} summarizes the attack, and Figure~\ref{fig:attack-overview} illustrates a representative execution.

\begin{algorithm}[t]
\caption{\textsc{Hush-Up} Attack Workflow}
\label{alg:hush-up}
\begin{algorithmic}[1]
\Require Assigned stage $\mathcal{M}_i$;
         received activation $\hat{h}_{i-1}$;
         calibrated budget $\rho_i$;
         synthetic prompt $S$;
         header $H$; threshold $\tau$
\Ensure Transmitted activation $\hat{h}_i$
         and evidence $e_i$

\State $h_i^f \gets \mathcal{M}_i(\hat{h}_{i-1})$
    \Comment{Faithful prefill computation}
\State $J_1(\delta) \gets \log p_\delta(\mathrm{EOS})$
\State $\delta_i \gets
    \Call{Optimize}{h_i^f,h_i^f,\rho_i,J_1}$

\If{$p_{\delta_i}(\mathrm{EOS}) < \tau$
    \textbf{and} the model generates $H$
    \textbf{and} $i<K$}
    \State $\tilde{h}_{i-1}^{H} \gets
        \Call{SimulateHeader}{S,H,i-1}$
        \Comment{Simulated header activations}
    \State $[\tilde{h}_i^{P};\tilde{h}_i^{H}] \gets
        \mathcal{M}_i([\hat{h}_{i-1};\tilde{h}_{i-1}^{H}])$
    \State $J_2(\delta) \gets
        \log \tilde{p}_\delta(\mathrm{EOS})$
        \Comment{Evaluated on
        $[\tilde{h}_i^{P}+\delta;\tilde{h}_i^{H}]$}
    \State $\delta_i \gets
        \Call{Optimize}{h_i^f,\tilde{h}_i^{P},\rho_i,J_2}$
        \Comment{Perturb prefill positions only}
\EndIf

\State $\hat{h}_i \gets h_i^f+\delta_i$
    \Comment{Apply to the native faithful activation}
\State $e_i \gets \Call{GenerateEvidence}{\hat{h}_i}$
\State Transmit $\hat{h}_i$ to the next worker
    and submit $e_i$ to the watchdog
\end{algorithmic}
\end{algorithm}

\subsection{Constraints and Objectives}
\label{sec:constraints-objectives}

\myparagraph{Benign Variation Profiling}
The adversary calibrates a stage-specific budget offline, independently of any inference request.
It collects a set $\mathcal{C}$ of prompts and, for each $x_j\in\mathcal{C}$, executes the model under two hardware environments $a$ and $b$ with identical weights, numerical precision, and inference configuration.
Using Equation~\eqref{eq:relative-distance}, it collects the discrepancies at the output of its stage,
\begin{equation}
    \mathcal{D}_i=\big\{\,d\big(h_i^{(a)}(x_j),h_i^{(b)}(x_j)\big) \mid x_j\in\mathcal{C}\,\big\},
\end{equation}
and sets the budget in Equation~\eqref{eq:attack-budget-constraint}
to their $q$-quantile:
\begin{equation}
    \rho_i = \operatorname{Quantile}_q(\mathcal{D}_i).
\end{equation}

Three considerations motivate this budget.
First, it is measured on faithful executions: deviations of this magnitude arise routinely across heterogeneous hardware, so a complete watchdog must accept them from honest workers.
Second, it is relative, which makes it comparable across prompts and stages whose activations differ widely in norm.
Third, it is calibrated separately for each stage boundary, since benign variation differs across depths.

\myparagraph{EOS Promoting}
Within the calibrated budget, the adversary seeks a
perturbation that induces immediate or near-immediate
termination. Let $p_\delta$ denote the next-token
distribution that the client obtains when $W_i$ transmits $h_i^f+\delta$ and all downstream stages execute faithfully.
The primary objective promotes EOS
as the first generated token:
\begin{equation}
    J_1(\delta)
    =
    \log p_\delta(\mathrm{EOS}).
\end{equation}

Some models, however, open every response with a model-specific header $H$, such as a reasoning-channel marker, that they generate autoregressively before any content (Figure~\ref{fig:attack-overview}).
Such a model may resist emitting EOS as its very first token, while being more susceptible to EOS promotion immediately after the header.
Targeting this position requires the header's activations at the adversary's input boundary, so that the header can be processed together with the received activations.
For the first worker, these are simply the header's embeddings, which do not depend on the prompt.
For a deeper worker, however, these activations depend on the original prompt through attention in the preceding stages, and thus cannot be reconstructed exactly without it.
\textsc{Hush-Up} therefore approximates them using a synthetic prompt $S$, built from the model's public chat template with a fixed generic system message (``You are a helpful AI assistant.'') and a user message padded so that $S$ has the same length as the observed prefill sequence.
The adversary appends $H$ to $S$, executes the stages preceding its own on this sequence, and extracts the activations $\tilde{h}_{i-1}^{H}$ at the header positions.

It then concatenates simulated header activations
with the received activations along the token dimension
and executes its stage:
\begin{equation}
    [\tilde{h}_i^{P};\tilde{h}_i^{H}]
    =
    \mathcal{M}_i
    ([\hat{h}_{i-1};\tilde{h}_{i-1}^{H}]),
\end{equation}
where $\tilde{h}_i^{P}$ denotes the outputs at the
original prefill positions in this augmented forward
pass. Under causal attention, these outputs equal $h_i^f$ in exact arithmetic, although
finite-precision execution may introduce small numerical
differences.

\begin{figure}[t]
    \centering
    \includegraphics[width=0.8\columnwidth]{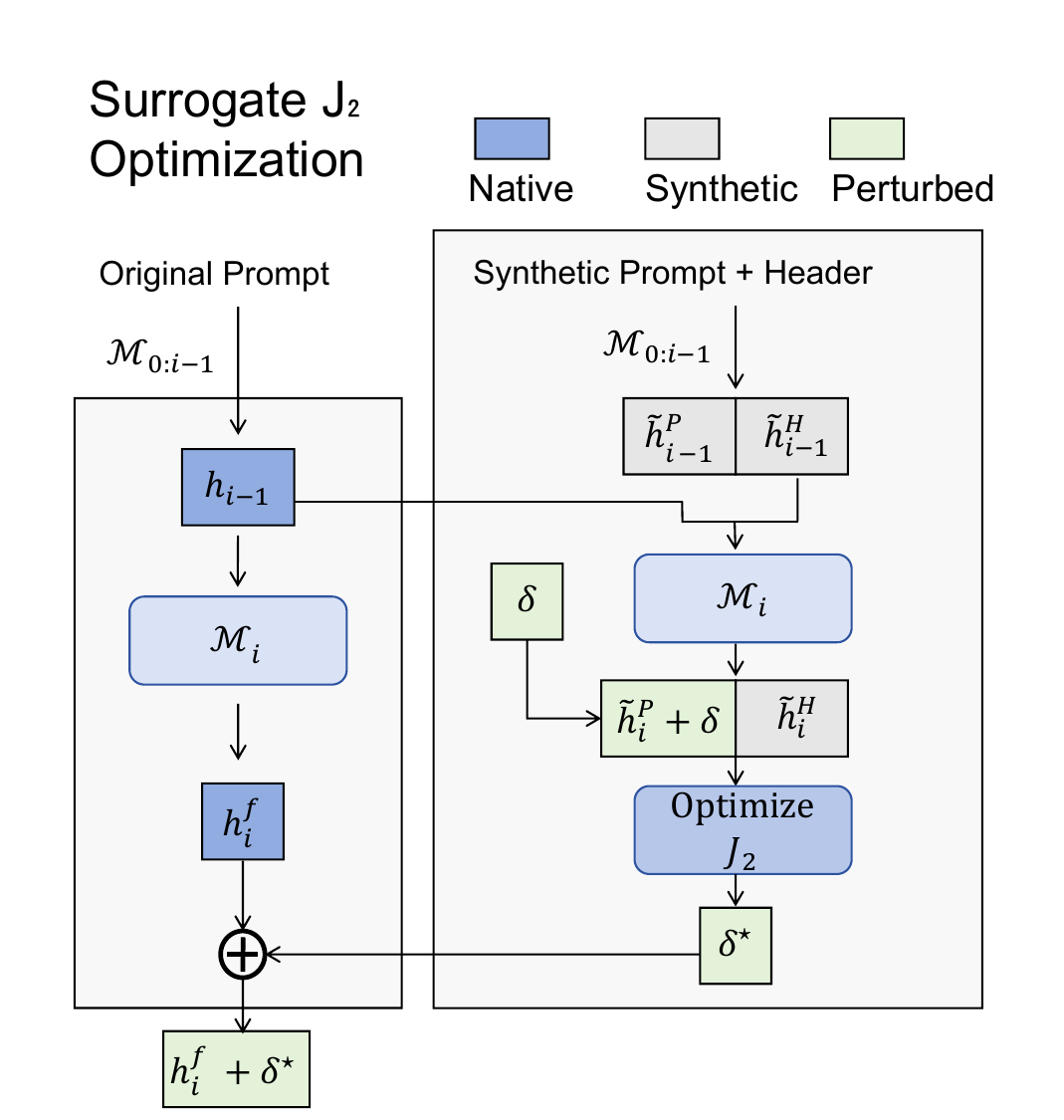}
    \vspace{-\baselineskip}
    \caption{Optimization of the surrogate objective $J_2$. Synthetic header
    activations are only used for perturbation optimization; the resulting
    perturbation is applied to the native prefill activation
    during distributed inference.}
    \label{fig:j2-optimization}
    \vspace{-\baselineskip}
\end{figure}

The surrogate objective evaluates the downstream
stages on
\begin{equation}
    [\tilde{h}_i^{P}+\delta;\tilde{h}_i^{H}]
\end{equation}
and promotes EOS after the appended header:
\begin{equation}
    J_2(\delta)
    =
    \log \tilde{p}_\delta(\mathrm{EOS}),
\end{equation}
where $\tilde{p}_\delta$ is the next-token distribution
at the end of the augmented sequence.
Only the original prefill positions are perturbed;
the synthetic header activations remain unperturbed.
Figure~\ref{fig:j2-optimization} illustrates the surrogate
$J_2$ optimization.

\subsection{Adversarial Perturbation Optimization}
\label{sec:optimization}

White-box access to the model allows the adversary
to evaluate the downstream stages and differentiate
the EOS objective with respect to the perturbation.
For $J_1$, gradients propagate from the first-token
prediction through the downstream stages to the
perturbed prefill activation.
For $J_2$, the appended header positions attend to the perturbed prefill positions in the downstream stages, which makes the post-header prediction depend on $\delta$ even though the header positions themselves are not perturbed.

We optimize unit-norm perturbation directions by
gradient ascent with backtracking, subject to
Equation~\eqref{eq:attack-budget-constraint}.
A realization operator $\mathcal{R}(h_i^f,u,\rho_i)$
constructs a feasible perturbation from a direction $u$,
ensuring that the transmitted activation satisfies
the prescribed budget.
For both objectives, the budget is enforced relative
to the faithful activation $h_i^f$, while the objective is evaluated on $h_i^f$ for $J_1$ and on $\tilde{h}_i^{P}$ for $J_2$.
Algorithm~\ref{alg:eos-optimizer} in
Appendix~\ref{app:optimizer} provides the full
optimization procedure.

The attack first optimizes $J_1$ to promote EOS as
the first generated token.
It switches to $J_2$ only when three conditions hold: the resulting EOS probability remains below $\tau$, the model generates $H$ autoregressively before its response, and at least one downstream stage remains (i.e., $i<K$).
Both objectives (i.e., $J_1$ and $J_2$) use the same perturbation budget,
and the perturbation is always restricted to the original
prefill positions.

The last condition matters at the final boundary ($i=K$), where only the output head remains downstream.
There, $J_1$ remains optimizable because the perturbation directly changes the hidden state used for the first-token prediction.
However, the output head operates position-wise and cannot propagate changes from the prefill positions to the unperturbed header positions, so $J_2$ does not depend on $\delta$.
The attack therefore uses only $J_1$ at this boundary.
\section{Evaluation}\label{sec:eval-setup}
\subsection{Experimental Setup}
\myparagraph{Models and Benchmarks} 
We evaluate three recent open-weight LLMs: Qwen3.8-27B~\cite{qwen38}, Muse-Glimmer-30B~\cite{museglimmerteam2026museglimmer30b}, and Gemma-4-31B-IT~\cite{gemmateam2026gemma4}. All models use BF16 weights and activations. We use each model's native inference template; reasoning configurations and generation-boundary details are provided in Appendix~\ref{app:experimental-details}.

We consider three benchmarks: GSM8K~\cite{cobbe2021trainingverifierssolvemath} for mathematical reasoning, HumanEval+~\cite{chen2021evaluatinglargelanguagemodels, liu2023codegeneratedchatgptreally} for code generation, and IFEval~\cite{zhou2023instructionfollowingevaluationlargelanguage} for instruction following. For each benchmark, we use a fixed set of 30 problems shared across all models and perturbation methods. GSM8K problems are drawn from its test split. We evaluate GSM8K by extracting the final numerical answer and comparing it against the ground truth, HumanEval+ using EvalPlus, and IFEval using the official strict prompt-level accuracy metric.

\myparagraph{Distributed Inference Setting} 
We partition each model's transformer blocks into five stages, each assigned to one worker.  
In the main experiments, the first worker is designated as malicious, while the others execute honestly. We further vary the malicious worker's position while keeping
the remaining setting fixed.

\begin{table*}[t]
\centering
\caption{Task degradation under \textsc{Hush-Up} (Ours), with the first worker as the malicious worker.
Each cell aggregates 120 trials.
Arrows indicate the attacker's preferred direction.
All values are percentages.}
\vspace{-.5\baselineskip}
\label{tab:task-degradation}
\small
\setlength{\tabcolsep}{6.1pt}
\begin{tabular}{llrrrrrrrrr}
\toprule
Model & Perturbation
& \multicolumn{3}{c}{GSM8K}
& \multicolumn{3}{c}{HumanEval+}
& \multicolumn{3}{c}{IFEval} \\
\cmidrule(lr){3-5}
\cmidrule(lr){6-8}
\cmidrule(lr){9-11}
& & Stop$\le$1 $\uparrow$ & Stop$\le$30 $\uparrow$ & Acc.\ $\downarrow$
  & Stop$\le$1 $\uparrow$ & Stop$\le$30 $\uparrow$ & Pass@1 $\downarrow$
  & Stop$\le$1 $\uparrow$ & Stop$\le$30 $\uparrow$ & Acc.\ $\downarrow$ \\
\midrule
Qwen3.8-27B & Clean
& 0 & 0 & 98.33 & 0 & 0 & 84.17 & 0 & 0 & 80 \\
& Gaussian
& 0 & 0 & 99.17 & 0 & 0 & 79.17 & 0 & 0 & 80 \\
& Uniform
& 0 & 0 & 98.33 & 0 & 0 & 84.17 & 0 & 0 & 83.33 \\
& \textbf{Ours}
& \textbf{100} & \textbf{100} & \textbf{0}
& \textbf{100} & \textbf{100} & \textbf{0}
& \textbf{100} & \textbf{100} & \textbf{0} \\
\midrule
Muse-Glimmer-30B & Clean
& 0 & 0 & 93.33 & 0 & 0 & 95 & 0 & 0 & 95.83 \\
& Gaussian
& 0 & 0 & 93.33 & 0 & 0 & 92.50 & 0 & 0 & 96.67 \\
& Uniform
& 0 & 0 & 93.33 & 0 & 0 & 92.50 & 0 & 0 & 95 \\
& \textbf{Ours}
& 0 & \textbf{96.67} & \textbf{3.33}
& \textbf{100} & \textbf{100} & \textbf{0}
& \textbf{96.67} & \textbf{96.67} & \textbf{3.33} \\
\midrule
Gemma-4-31B-IT & Clean
& 0 & 0 & 92.50 & 0 & 0 & 70.83 & 0 & 0 & 85 \\
& Gaussian
& 0 & 0 & 93.33 & 0 & 0 & 75.83 & 0 & 0 & 85 \\
& Uniform
& 0 & 0 & 91.67 & 0 & 0 & 70 & 0 & 0 & 84.17 \\
& \textbf{Ours}
& \textbf{34.17} & \textbf{79.17} & \textbf{7.50}
& \textbf{76.67} & \textbf{90} & \textbf{2.50}
& \textbf{16.67} & \textbf{68.33} & \textbf{10} \\
\bottomrule
\end{tabular}
\vspace{-.5\baselineskip}
\end{table*}
\begin{table*}[t]
\centering
\caption{Watchdog outcomes under \textsc{Hush-Up} across three benchmarks, with the first worker as the malicious worker.
DF: detection failure; AS: attribution success; AF: attribution failure.
All values are percentages.}
\vspace{-.5\baselineskip}
\label{tab:verification-outcomes}
\small
\setlength{\tabcolsep}{8.9pt}
\begin{tabular}{lrrrrrrrrrrrr}
\toprule
& \multicolumn{3}{c}{TopLoc}
& \multicolumn{3}{c}{SENTINEL}
& \multicolumn{3}{c}{Activation-DiFR}
& \multicolumn{3}{c}{TAO} \\
\cmidrule(lr){2-4}
\cmidrule(lr){5-7}
\cmidrule(lr){8-10}
\cmidrule(lr){11-13}
Model
& DF $\uparrow$ & AS $\downarrow$ & AF $\uparrow$
& DF $\uparrow$ & AS $\downarrow$ & AF $\uparrow$
& DF $\uparrow$ & AS $\downarrow$ & AF $\uparrow$
& DF $\uparrow$ & AS $\downarrow$ & AF $\uparrow$ \\
\midrule
Qwen3.8-27B
& 52.22 & 0.00 & 47.78
& 4.44 & 0.00 & 95.56
& 1.11 & 0.00 & 98.89
& 4.44 & 0.00 & 95.56 \\
Muse-Glimmer-30B
& 100.00 & 0.00 & 0.00
& 97.78 & 0.00 & 2.22
& 1.11 & 0.00 & 98.89
& 52.22 & 0.00 & 47.78 \\
Gemma-4-31B-IT
& 0.00 & 0.00 & 100.00
& 32.22 & 5.56 & 62.22
& 0.00 & 0.00 & 100.00
& 0.00 & 100.00 & 0.00 \\
\bottomrule
\end{tabular}
\vspace{-\baselineskip}
\end{table*}

\myparagraph{Attack Configuration} 
We conduct our main experiments on NVIDIA A100 GPUs and use H100 for profiling benign cross-hardware numerical variation. We further evaluate hardware generalization and transferability using A100, A6000, and L40 GPUs.

For each model and injection boundary, we calibrate $\rho_i$ at the 90th percentile of benign distances using 160 inputs, with 40 additional inputs held out
for validation. Model weights, numerical precision, and inference configuration are held constant across environments. At the first worker boundary (20\% depth), the resulting budgets are $\rho=0.4646\%$ for Qwen, $0.8813\%$ for Muse, and $0.3767\%$ for Gemma. These budgets remain fixed across prompts and optimization objectives.
We optimize perturbations in FP32 and enforce the budget on the activation difference realized after BF16 rounding. Detailed configurations are provided in Appendix~\ref{app:experimental-details}.

\myparagraph{Watchdogs}

We evaluate four recent watchdogs.
\textit{TopLoc}~\cite{ong2025toploc} constructs locality-sensitive fingerprints of intermediate activations and compares them with those obtained from reference computations. 
\textit{Activation-DiFR}~\cite{karvonen2025difr} projects activations into lower-dimensional representations and detects discrepancies from trusted references using calibrated detectors. 
\textit{TAO}~\cite{yao2026tao} verifies intermediate computations against operator-level tolerance bounds, with discrepancies handled through recursive localization.
\textit{SENTINEL}~\cite{dolatabadi2026sentinel} compares activations against historical exponential moving average (EMA) references using multiple distance metrics and flags deviations exceeding calibrated thresholds. 
In our experiments, we replace pipeline localization with an oracle-assisted attribution test. This gives the watchdogs additional advantages by removing localization uncertainty while retaining each watchdog's discrepancy test. We collect paired clean and perturbed activations and use the clean activations as verification references when reference re-execution is required. Implementation details appear in
Appendix~\ref{app:experimental-details}.

\myparagraph{Metrics}

We measure premature termination
using Stop$\le$1 and Stop$\le$30: the fraction of generation trials that terminate within the first 1 and 30 generated tokens, respectively. Token counts begin at the native generation boundary, including generated header tokens and the terminating token. 
We measure task performance using answer accuracy for GSM8K, pass@1 for HumanEval+, and strict prompt-level accuracy for IFEval. All task and termination metrics are averaged over the evaluated problems and decoding trials.

To measure whether \textsc{Hush-Up} can evade watchdog identification, we report the three mutually exclusive outcomes of Section~\ref{sec:problem}: detection failure (DF), when the watchdog accepts the attacked execution; attribution success (AS), when it rejects the execution and also 
rejects at the oracle-provided malicious worker's location;
and attribution failure (AF), when it rejects the execution without successful attribution, either by identifying an honest worker as responsible, or by failing to identify any responsible worker.

\subsection{Experimental Results}\label{sec:results}

\myparagraph{Attack Effectiveness}
We compare \textsc{Hush-Up} against honest execution (Clean) and two random perturbation baselines, Gaussian and Uniform. These random perturbations are sampled element-wise from a standard normal distribution and from $\mathcal{U}[-1,1]$, respectively, and scaled to match the same relative perturbation budget as \textsc{Hush-Up}.
We perform four independently seeded decoding trials per problem, yielding 120 trials per model--benchmark--method combination. For \textsc{Hush-Up}, the optimized perturbation is reused across the four trials; random baselines use independently sampled perturbation directions for each trial. 

\textit{\textsc{Hush-Up} effectively halts generation across models and tasks, whereas same-budget random perturbations largely preserve normal behavior.}
Table~\ref{tab:task-degradation} reports premature termination rates and task performance.
Task success falls from $70.83$--$98.33\%$ under clean execution to at most $10\%$ across all nine attacked model--benchmark combinations, reaching zero in four.
Neither random baseline produces termination within 30 tokens. This contrast shows that perturbation magnitude alone does not determine behavioral impact; the direction
of the perturbation is critical.

Qwen terminates at the first generated token across all three benchmarks.
Muse exhibits predominantly immediate termination on HumanEval+ and IFEval, while on GSM8K it terminates immediately after generating its protocol header. Gemma exhibits more varied behavior, but still terminates within 30 tokens in most trials.
Moreover, we observe nonsensical outputs, such as repetitive text, showing that \textsc{Hush-Up}'s task-level impact extends beyond premature termination (Appendix~\ref{app:qualitative-examples}).

\myparagraph{Watchdog Effectiveness}
We further evaluate the effectiveness of watchdogs under our attack.
Table~\ref{tab:verification-outcomes} reports detection and attribution outcomes under our oracle-assisted evaluation protocol for four watchdogs.
\textit{\textsc{Hush-Up} effectively evades attribution, even when the watchdog has oracle knowledge of the adversary's position.}
Attribution success is zero in ten of the twelve configurations.
Complete results are deferred to Appendix~\ref{app:verification-outcomes}.

TopLoc accepts all attacked Muse executions and over half of attacked Qwen executions.
Although it rejects every attacked Gemma execution, none is correctly attributed.
Activation-DiFR rejects almost all attacked executions across the three models, yet achieves zero attribution success.
Thus, execution rejection does not ensure correct
attribution.
SENTINEL accepts $97.78\%$ of attacked Muse executions,
matching its clean acceptance rate.
For Qwen and Gemma, its clean acceptance rates are
only $12.22\%$ and $28.89\%$, respectively, so its
high rejection rates under attack must be interpreted
alongside substantial false-alarm rates.
Its attribution success for Gemma is only $5.56\%$,
matching that under random perturbations.

TAO achieves zero attribution success for Qwen and Muse.
Gemma--TAO is the sole configuration with $100\%$ attribution success in Table~\ref{tab:verification-outcomes}, which is analyzed in Appendix~\ref{app:verification-outcomes}.
However, this does not extend uniformly across injection depths (Table~\ref{tab:injection-depth}).

\begin{table}[t]
\centering
\caption{Effect of attacker configuration on GSM8K accuracy
and TAO attribution success (AS). All values are percentages.}
\label{tab:injection-depth}
\vspace{-.5\baselineskip}
\small
\setlength{\tabcolsep}{6.7pt}
\begin{tabular}{lrrrrrr}
\toprule
& \multicolumn{2}{c}{Qwen3.8-27B}
& \multicolumn{2}{c}{Muse-Glimmer-30B}
& \multicolumn{2}{c}{Gemma-4-31B-IT} \\
\cmidrule(lr){2-3}
\cmidrule(lr){4-5}
\cmidrule(l){6-7}
Depth
& Acc. $\downarrow$ & AS $\downarrow$
& Acc. $\downarrow$ & AS $\downarrow$
& Acc. $\downarrow$ & AS $\downarrow$\\
\midrule
20\%  & 0.00 & 0.00 & 3.33 & 0.00 & 6.67 & 100.00 \\
40\%  & 0.00 & 0.00 & 0.00 & 0.00 & 0.00 & 46.67 \\
60\%  & 0.00 & 0.00 & 0.00 & 0.00 & 0.00 & 30.00 \\
80\%  & 0.00 & 0.00 & 0.00 & 0.00 & 0.00 & 80.00 \\
100\% & 0.00 & 0.00 & 0.00 & 0.00 & 0.00 & 100.00 \\
\bottomrule
\end{tabular}
\vspace{-\baselineskip}
\end{table}

\myparagraph{Effects of Attacker Configuration}
To examine \textsc{Hush-Up}'s effectiveness across malicious-worker positions, we vary the injection boundary from $20\%$ to $100\%$ of model depth, corresponding to the output boundaries of the five workers. At $100\%$ depth, the injection and final boundaries coincide.
For each boundary, we independently calibrate the perturbation budget and optimize the perturbation, while fixing the problems and optimization
configuration. Each setting uses 30 GSM8K problems with one decoding
trial per problem.

\begin{table}[t]
\centering
\caption{Hardware generalization on GSM8K.
Perturbations are independently constructed and evaluated on each hardware configuration. Each model uses the same 30 problems with one decoding trial per problem. All values are percentages.}
\vspace{-.5\baselineskip}
\label{tab:hardware-generalization}
\small
\setlength{\tabcolsep}{4.4pt}
\begin{tabular}{llrrr}
\toprule
Model & Hardware & Stop$\le$1 $\uparrow$ & Stop$\le$30 $\uparrow$ & Acc.\ $\downarrow$ \\
\midrule
Qwen3.8-27B & A100 & 100.00 & 100.00 & 0.00 \\
 & $2\times$ L40 & 100.00 & 100.00 & 0.00 \\
 & $2\times$ RTX A6000 & 100.00 & 100.00 & 0.00 \\
\midrule
Muse-Glimmer-30B & A100 & 0.00 & 96.67 & 3.33 \\
 & $2\times$ L40 & 16.67 & 100.00 & 0.00 \\
 & $2\times$ RTX A6000 & 10.00 & 100.00 & 0.00 \\
\midrule
Gemma-4-31B-IT & A100 & 50.00 & 90.00 & 3.33 \\
 & $2\times$ L40 & 46.67 & 96.67 & 0.00 \\
 & $2\times$ RTX A6000 & 53.33 & 86.67 & 3.33 \\
\bottomrule
\end{tabular}
\vspace{-.5\baselineskip}
\end{table}

\begin{table}[t]
\centering
\captionsetup{skip=3pt}
\caption{Cross-hardware transferability on GSM8K.
Each cell reports Stop$\le$30 / Acc.\ (\%).
Rows and columns denote source and target hardware,
respectively.
A6000 and L40 use dual-GPU configurations.}
\label{tab:hardware-transfer}
\small
\setlength{\tabcolsep}{3pt}
\begin{tabular*}{\columnwidth}{
    @{\extracolsep{\fill}}llrrr@{}
}
\toprule
& & \multicolumn{3}{c}{Target} \\
\cmidrule(lr){3-5}
Model & Source & A100 & A6000 & L40 \\
\midrule
Qwen
& A100  & 100.00 / 0.00 & 100.00 / 0.00 & 100.00 / 0.00 \\
& A6000 & 100.00 / 0.00 & 100.00 / 0.00 & 100.00 / 0.00 \\
& L40   & 100.00 / 0.00 & 100.00 / 0.00 & 100.00 / 0.00 \\
\midrule
Muse
& A100  & 96.67 / 3.33 & 96.67 / 3.33 & 96.67 / 3.33 \\
& A6000 & 100.00 / 0.00 & 100.00 / 0.00 & 100.00 / 0.00 \\
& L40   & 100.00 / 0.00 & 100.00 / 0.00 & 100.00 / 0.00 \\
\midrule
Gemma
& A100  & 90.00 / 3.33 & 90.00 / 3.33 & 90.00 / 3.33 \\
& A6000 & 86.67 / 3.33 & 86.67 / 3.33 & 86.67 / 3.33 \\
& L40   & 96.67 / 0.00 & 96.67 / 3.33 & 96.67 / 0.00 \\
\bottomrule
\end{tabular*}
\vspace{-4pt}
\end{table}

\textit{\textsc{Hush-Up} remains effective across malicious-worker positions, including the final worker boundary.}
As reported in Table~\ref{tab:injection-depth},
task accuracy remains at or below $6.67\%$ across all injection depths, showing that the attack remains effective regardless of the malicious worker's position.
Moreover, $J_1$ alone suffices at every boundary beyond the first worker, without switching to $J_2$. We therefore additionally evaluate $J_2$ at these boundaries
in Appendix~\ref{app:j2-analysis}.

Attribution failures also persist across worker positions.
TAO achieves zero attribution success for Qwen and Muse at every depth, as does TopLoc for all three models (Appendix~\ref{app:injection-depth}).
For Gemma, TAO's attribution success varies non-monotonically with depth, showing that its perfect attribution at the first worker boundary does not extend uniformly to other positions.

\myparagraph{Effects of Hardware Configuration}
To examine whether \textsc{Hush-Up} remains effective across hardware configurations and whether perturbations constructed on one configuration transfer to another, we evaluate all three models on the same 30 GSM8K problems, using one decoding trial per problem. We consider three hardware configurations: a single NVIDIA A100, two NVIDIA L40 GPUs, and two NVIDIA RTX A6000 GPUs. For hardware generalization, we independently calibrate the perturbation budget against the H100 reference and construct perturbations on each configuration. For cross-hardware transferability, we evaluate all
source--target combinations, reusing the source perturbations without re-optimization.

Our results show that \textit{\textsc{Hush-Up} remains effective across
hardware configurations, and its perturbations
transfer without re-optimization.}
Table~\ref{tab:hardware-generalization} reports consistent
generation suppression across all three models and
hardware configurations, while Table~\ref{tab:hardware-transfer}
demonstrates that transferred perturbations retain effectiveness
comparable to their source-hardware performance.
\section{Discussion}\label{sec:discussions}
\myparagraph{Limitations of Local Verification}
Our findings highlight a gap between local numerical verification and end-to-end accountability. A perturbation that remains within a watchdog's tolerance at its source can propagate through subsequent, faithfully executed stages and cause
substantial task failures. Although the resulting deviations may eventually trigger detection, the worker at which a discrepancy becomes observable need not be the one that
introduced it.
Consequently, detecting an anomalous execution does not necessarily imply correctly attributing it to the malicious worker.

\myparagraph{Possible Mitigation}
One mitigation is to verify local computations using
their actual inputs rather than independently
reconstructed reference inputs~\cite{
wang2026verillmlightweightframeworkpublicly,yao2026tao}.
By accounting for perturbations introduced upstream, this approach may reduce false attribution to honest downstream workers. However, it does not by itself identify the malicious worker if the injected perturbation remains within the watchdog's tolerance at its source. Moreover, retaining full boundary activations for replay
introduces storage overhead relative to compact activation fingerprints~\cite{ong2025toploc}.

\myparagraph{Scope and Limitations}
Our study focuses on premature termination through
EOS promotion; we do not systematically evaluate adversarial objectives
targeting other forms of task failure.
Although our perturbation budgets are calibrated against benign cross-hardware numerical variation, they constrain the magnitude of deviations rather than their
distributional structure. Our experiments primarily use BF16 inference;
other numerical precisions and quantization schemes may exhibit different numerical variation and attack opportunities.

\section{Related Work}

\myparagraph{Distributed LLM Inference Frameworks}
LLMs impose substantial memory and computational demands, making local deployment difficult on consumer-grade GPUs.
Offloading model weights to CPU memory or disk can incur substantial inference
latency~\cite{pmlr-v202-sheng23a,
song2024powerinferfastlargelanguage},
while commercial inference APIs typically provide limited access to model internals~\cite{borzunov2023petals}.
Pipeline-parallel distributed LLM inference alleviates these challenges by partitioning models into consecutive stages and assigning each stage to a participating worker.
Petals~\cite{borzunov2023petals,borzunov2023distributed}
distributes transformer blocks across collaborative servers,
while LinguaLinked~\cite{zhao2023lingualinkeddistributedlargelanguage}
enables mobile inference using ring-structured communication.
HexGen~\cite{pmlr-v235-jiang24f} supports asymmetric
partitioning across heterogeneous GPUs, with
HexGen-2~\cite{ICLR2025_0b941a1e} and
HexGen-3~\cite{jiang2026hexgen} introducing
prefill--decoding and attention--FFN disaggregation,
respectively.
Parallax~\cite{tong2025parallaxefficientllminference}
optimizes model allocation and request-time pipeline selection
across heterogeneous GPUs.

These frameworks focus on pooling computing resources across participating devices for efficient LLM inference, while the integrity of those contributions remains a separate challenge. Our work instead investigates how malicious workers can induce task failures.

\myparagraph{Attacks against Distributed Inference}
Workers in distributed inference have direct access to intermediate activations.
Prior attacks have shown that this access can enable reconstruction of private prompts~\cite{luo2025promptinferenceattackdistributed,qu2025promptinversionattackcollaborative}. This access also raises concerns beyond privacy, since modifying LLM activations during inference can steer model outputs and behaviors~\cite{ fort2023scalinglawsadversarialattacks,lee2025programmingrefusalconditionalactivation,10.1145/3696410.3714640}.
ActivationBackdoor~\cite{10.1145/3770855.3818136} demonstrates inference-time integrity attacks through conditional activation steering. However, its optimization does not explicitly minimize perturbation magnitude, while its detection evaluation focuses on
statistical activation anomalies through BadActs~\cite{yi-etal-2024-badacts}, rather than verification of computational correctness.

In contrast, we study perturbations constrained to a benign
numerical scale and evaluate their impact on task performance,
as well as their detection and attribution by
execution-integrity watchdogs.

\myparagraph{Defenses}
Cryptographic verification provides strong correctness
guarantees~\cite{sun2024zkllm,qu2025zkgpt,
kang2022scalingtrustlessdnninference}, but incurs
substantial overhead, limiting its applicability to
large-scale LLM inference~\cite{ong2025toploc}.

Lighter-weight approaches offer an alternative by checking intermediate activations or evidence derived from them.
SVIP~\cite{sun2026svipverifiableinferenceopensource} uses a secret-based protocol built on processed hidden representations to detect substitution of the requested model.
TopLoc~\cite{ong2025toploc} constructs locality-sensitive
fingerprints of intermediate activations for efficient verification while tolerating benign numerical variation. TopLoc v2~\cite{primeintellect2025synthetic2} extends
TopLoc to distributed inference, using stage-by-stage replay upon verification failure to identify the first faulty worker.
TAO~\cite{yao2026tao} combines theoretical floating-point error bounds with empirically calibrated tolerances and employs an interactive dispute protocol to localize incorrect computations.
Activation-DiFR~\cite{karvonen2025difr} uses random orthogonal projections to compress intermediate activations into fingerprints for comparison with trusted reference computations.
SENTINEL~\cite{dolatabadi2026sentinel} uses EMA-based activation monitoring for stagewise integrity verification in decentralized training.
VeriLLM~\cite{wang2026verillmlightweightframeworkpublicly} uses game-theoretic incentives to support decentralized LLM inference verification without relying on a
single trusted verifier.

Building on prior efforts toward fault localization and attribution~\cite{primeintellect2025synthetic2,yao2026tao,dolatabadi2026sentinel,
cihangiroglu2026integritypeertopeerdistributedllm}, we examine whether these mechanisms can correctly attribute deviations induced by benign-scale adversarial perturbations.
Our evaluation distinguishes detecting an anomalous execution from identifying the malicious worker that introduces the deviation.

\section{Conclusion}
We presented \textsc{Hush-Up}, which induces immediate or near-immediate generation termination through constrained activation perturbations in distributed LLM inference.

Across multiple model families and hardware configurations, these optimized perturbations at the scale of benign numerical variation substantially degrade task performance.
Meanwhile, existing activation-based execution-integrity watchdogs often accept attacked executions or reject them without correctly attributing the deviation to the malicious worker, even with oracle knowledge of the injection boundary.
These findings expose a gap between local activation verification and end-to-end accountability: the numerical tolerance needed to accommodate honest executions creates an attack surface that malicious workers can exploit to evade attribution. Existing activation-based watchdogs therefore do not reliably ensure accountability for adversarial deviations in distributed LLM inference.

\bibliographystyle{ACM-Reference-Format}
\bibliography{ref}

@misc{zamir2026noncomposability,
  title        = {A Note on Non-Composability of Layerwise Approximate Verification for Neural Inference},
  author       = {Zamir, Or},
  year         = {2026},
  eprint       = {2602.15756},
  archivePrefix = {arXiv},
  doi          = {10.48550/arXiv.2602.15756},
  url          = {https://arxiv.org/abs/2602.15756}
}

@inproceedings{borzunov2023petals,
  title={Petals: Collaborative inference and fine-tuning of large models},
  author={Borzunov, Alexander and Baranchuk, Dmitry and Dettmers, Tim and Riabinin, Maksim and Belkada, Younes and Chumachenko, Artem and Samygin, Pavel and Raffel, Colin},
  booktitle={Proceedings of the 61st Annual Meeting of the Association for Computational Linguistics (Volume 3: System Demonstrations)},
  pages={558--568},
  year={2023}
}

@article{borzunov2023distributed,
  title={Distributed inference and fine-tuning of large language models over the internet},
  author={Borzunov, Alexander and Ryabinin, Max and Chumachenko, Artem and Baranchuk, Dmitry and Dettmers, Tim and Belkada, Younes and Samygin, Pavel and Raffel, Colin},
  journal={Advances in neural information processing systems},
  volume={36},
  pages={12312--12331},
  year={2023}
}

@inproceedings{sun2024zkllm,
  title={zkllm: Zero knowledge proofs for large language models},
  author={Sun, Haochen and Li, Jason and Zhang, Hongyang},
  booktitle={Proceedings of the 2024 on ACM SIGSAC Conference on Computer and Communications Security},
  pages={4405--4419},
  year={2024}
}

@inproceedings{qu2025zkgpt,
  title={{zkGPT}: An Efficient Non-interactive Zero-knowledge Proof Framework for {LLM} Inference},
  author={Qu, Wenjie and Sun, Yijun and Liu, Xuanming and Lu, Tao and Guo, Yanpei and Chen, Kai and Zhang, Jiaheng},
  booktitle={34th USENIX Security Symposium (USENIX Security 25)},
  pages={2045--2063},
  year={2025}
}

@inproceedings{
ong2025toploc,
title={{TOPLOC}: A Locality Sensitive Hashing Scheme for Trustless Verifiable Inference},
author={Jack Min Ong and Matthew Di Ferrante and Aaron Pazdera and Ryan Garner and Sami Jaghouar and Manveer Basra and Max Ryabinin and Johannes Hagemann},
booktitle={Forty-second International Conference on Machine Learning},
year={2025},
}

@article{karvonen2025difr,
  title={DiFR: Inference verification despite nondeterminism},
  author={Karvonen, Adam and Reuter, Daniel and Rinberg, Roy and Marks, Luke and Garriga-Alonso, Adri{\`a} and Warr, Keri},
  journal={arXiv preprint arXiv:2511.20621},
  year={2025}
}

@inproceedings{yao2026tao,
  title={{TAO}: Tolerance-Aware Optimistic Verification for Floating-Point Neural Networks},
  author={Yao, Jianzhu and Su, Hongxu and Liao, Taobo and Cheng, Zerui and Zhang, Huan and Wang, Xuechao and Viswanath, Pramod},
  booktitle={Proceedings of the 21st European Conference on Computer Systems},
  pages={1515--1532},
  year={2026}
}

@article{dolatabadi2026sentinel,
  title={SENTINEL: Stagewise Integrity Verification for Pipeline Parallel Decentralized Training},
  author={Dolatabadi, Hadi Mohaghegh and Ajanthan, Thalaiyasingam and Ramasinghe, Sameera and Koneputugodage, Chamin P Hewa and Avraham, Gil and Zuo, Yan and Shevchenko, Violetta and Long, Alexander},
  journal={arXiv preprint arXiv:2603.03592},
  year={2026}
}

@inproceedings{
yuan2025understanding,
title={Understanding and Mitigating Numerical Sources of Nondeterminism in {LLM} Inference},
author={Jiayi Yuan and Hao Li and Xinheng Ding and Wenya Xie and Yu-Jhe Li and Wentian Zhao and Kun Wan and Jing Shi and Xia Hu and Zirui Liu},
booktitle={The Thirty-ninth Annual Conference on Neural Information Processing Systems},
year={2025},
}

@misc{tong2025parallaxefficientllminference,
      title={Parallax: Efficient LLM Inference Service over Decentralized Environment}, 
      author={Chris Tong and Youhe Jiang and Gufeng Chen and Tianyi Zhao and Sibian Lu and Wenjie Qu and Eric Yang and Lynn Ai and Binhang Yuan},
      year={2025},
      eprint={2509.26182},
      archivePrefix={arXiv},
      primaryClass={cs.DC},
      url={https://arxiv.org/abs/2509.26182}, 
}

@misc{chen2026distributedgenerativeinferencellm,
      title={Distributed Generative Inference of LLM at Internet Scales with Multi-Dimensional Communication Optimization}, 
      author={Jiu Chen and Shuangyan Yang and Xu Xiong and Hexiao Duan and Xinran Zhang and Jie Ren and Dong Li},
      year={2026},
      eprint={2604.21072},
      archivePrefix={arXiv},
      primaryClass={cs.DC},
      url={https://arxiv.org/abs/2604.21072}, 
}

@InProceedings{pmlr-v235-jiang24f,
  title = 	 {{H}ex{G}en: Generative Inference of Large Language Model over Heterogeneous Environment},
  author =       {Jiang, Youhe and Yan, Ran and Yao, Xiaozhe and Zhou, Yang and Chen, Beidi and Yuan, Binhang},
  booktitle = 	 {Proceedings of the 41st International Conference on Machine Learning},
  pages = 	 {21946--21961},
  year = 	 {2024},
  editor = 	 {Salakhutdinov, Ruslan and Kolter, Zico and Heller, Katherine and Weller, Adrian and Oliver, Nuria and Scarlett, Jonathan and Berkenkamp, Felix},
  volume = 	 {235},
  series = 	 {Proceedings of Machine Learning Research},
  month = 	 {21--27 Jul},
  publisher =    {PMLR},
  url = 	 {https://proceedings.mlr.press/v235/jiang24f.html}
}

@misc{wu2025deserveaffordableofflinellm,
      title={DeServe: Towards Affordable Offline LLM Inference via Decentralization}, 
      author={Linyu Wu and Xiaoyuan Liu and Tianneng Shi and Zhe Ye and Dawn Song},
      year={2025},
      eprint={2501.14784},
      archivePrefix={arXiv},
      primaryClass={cs.DC},
      url={https://arxiv.org/abs/2501.14784}, 
}

@misc{cihangiroglu2026integritypeertopeerdistributedllm,
      title={Integrity of peer-to-peer distributed LLM inference under malicious nodes}, 
      author={Mert Cihangiroglu and Antonino Nocera},
      year={2026},
      eprint={2607.19490},
      archivePrefix={arXiv},
      primaryClass={cs.CR},
      url={https://arxiv.org/abs/2607.19490}, 
}

@inproceedings{10.1145/3627703.3650088,
author = {Chen, Bing-Jyue and Waiwitlikhit, Suppakit and Stoica, Ion and Kang, Daniel},
title = {ZKML: An Optimizing System for ML Inference in Zero-Knowledge Proofs},
year = {2024},
isbn = {9798400704376},
publisher = {Association for Computing Machinery},
address = {New York, NY, USA},
url = {https://doi.org/10.1145/3627703.3650088},
doi = {10.1145/3627703.3650088},
booktitle = {Proceedings of the Nineteenth European Conference on Computer Systems},
pages = {560–574},
numpages = {15},
location = {Athens, Greece},
series = {EuroSys '24}
}

@misc{kang2022scalingtrustlessdnninference,
      title={Scaling up Trustless DNN Inference with Zero-Knowledge Proofs}, 
      author={Daniel Kang and Tatsunori Hashimoto and Ion Stoica and Yi Sun},
      year={2022},
      eprint={2210.08674},
      archivePrefix={arXiv},
      primaryClass={cs.CR},
      url={https://arxiv.org/abs/2210.08674}, 
}

@inproceedings{NIPS2017_6048ff4e,
 author = {Ghodsi, Zahra and Gu, Tianyu and Garg, Siddharth},
 booktitle = {Advances in Neural Information Processing Systems},
 editor = {I. Guyon and U. Von Luxburg and S. Bengio and H. Wallach and R. Fergus and S. Vishwanathan and R. Garnett},
 pages = {},
 publisher = {Curran Associates, Inc.},
 title = {SafetyNets: Verifiable Execution of Deep Neural Networks on an Untrusted Cloud},
 url = {https://proceedings.neurips.cc/paper_files/paper/2017/file/6048ff4e8cb07aa60b6777b6f7384d52-Paper.pdf},
 volume = {30},
 year = {2017}
}

@misc{nvidia2023tensorrtllm,
  author = {{NVIDIA}},
  title = {{NVIDIA TensorRT-LLM Supercharges Large Language Model Inference on NVIDIA H100 GPUs}},
  year = {2023},
  howpublished = {NVIDIA Technical Blog},
  url = {https://developer.nvidia.com/blog/nvidia-tensorrt-llm-supercharges-large-language-model-inference-on-nvidia-h100-gpus/}
}

@misc{primeintellect2025synthetic2,
  author       = {{Prime Intellect}},
  title        = {{SYNTHETIC-2 Release: Four Million Collaboratively Generated Reasoning Traces}},
  year         = {2025},
  month        = jul,
  day          = {10},
  howpublished = {\url{https://www.primeintellect.ai/blog/synthetic-2-release}},
  note         = {Official blog post}
}

@inproceedings{ICLR2025_0b941a1e,
 author = {JIANG, YOUHE and Yan, Ran and Yuan, Binhang},
 booktitle = {International Conference on Learning Representations},
 editor = {Y. Yue and A. Garg and N. Peng and F. Sha and R. Yu},
 pages = {3985--4002},
 title = {HexGen-2: Disaggregated Generative Inference of LLMs in Heterogeneous Environment},
 url = {https://proceedings.iclr.cc/paper_files/paper/2025/file/0b941a1e5fbce23fe46b049999d04ed0-Paper-Conference.pdf},
 volume = {2025},
 year = {2025}
}

@inproceedings{
jiang2026hexgen,
title={HexGen-3: A Fully Disaggregated {LLM} Serving Framework with Fine-Grained Heterogeneous Resource Autoscaling},
author={Youhe Jiang and Wenshuang Li and You Peng and Jintao Zhang and Ran Yan and Jianfei Chen and Xu Han and Fangcheng Fu and Binhang Yuan},
booktitle={Forty-third International Conference on Machine Learning},
year={2026},
url={https://openreview.net/forum?id=fVG57FKBNp}
}

@misc{luo2025promptinferenceattackdistributed,
      title={Prompt Inference Attack on Distributed Large Language Model Inference Frameworks}, 
      author={Xinjian Luo and Ting Yu and Xiaokui Xiao},
      year={2025},
      eprint={2503.09291},
      archivePrefix={arXiv},
      primaryClass={cs.CR},
      url={https://arxiv.org/abs/2503.09291}, 
}

@misc{qu2025promptinversionattackcollaborative,
      title={Prompt Inversion Attack against Collaborative Inference of Large Language Models}, 
      author={Wenjie Qu and Yuguang Zhou and Yongji Wu and Tingsong Xiao and Binhang Yuan and Yiming Li and Jiaheng Zhang},
      year={2025},
      eprint={2503.09022},
      archivePrefix={arXiv},
      primaryClass={cs.CR},
      url={https://arxiv.org/abs/2503.09022}, 
}

@inproceedings{10.1145/3770855.3818136,
author = {Su, Zichun and Zhang, Mi and Zhang, Xiaohan and Hong, Geng and You, Xiaoyu and Yang, Min},
title = {ActivationBackdoor: Backdooring Large Language Models in Collaborative Inference via Intermediate Activations},
year = {2026},
isbn = {9798400722592},
publisher = {Association for Computing Machinery},
address = {New York, NY, USA},
url = {https://doi.org/10.1145/3770855.3818136},
doi = {10.1145/3770855.3818136},
booktitle = {Proceedings of the 32nd ACM SIGKDD Conference on Knowledge Discovery and Data Mining V.2},
pages = {4429–4440},
numpages = {12},
location = {Jeju Island, Republic of Korea},
series = {KDD '26}
}

@inproceedings{yi-etal-2024-badacts,
    title = "{B}ad{A}cts: A Universal Backdoor Defense in the Activation Space",
    author = "Yi, Biao  and
      Chen, Sishuo  and
      Li, Yiming  and
      Li, Tong  and
      Zhang, Baolei  and
      Liu, Zheli",
    editor = "Ku, Lun-Wei  and
      Martins, Andre  and
      Srikumar, Vivek",
    booktitle = "Findings of the {A}ssociation for {C}omputational {L}inguistics: {ACL} 2024",
    month = aug,
    year = "2024",
    address = "Bangkok, Thailand",
    publisher = "Association for Computational Linguistics",
    url = "https://aclanthology.org/2024.findings-acl.317/",
    doi = "10.18653/v1/2024.findings-acl.317",
    pages = "5339--5352"
}

@misc{sun2026svipverifiableinferenceopensource,
      title={SVIP: Towards Verifiable Inference of Open-source Large Language Models}, 
      author={Yifan Sun and Yuhang Li and Yue Zhang and Yuchen Jin and Huan Zhang},
      year={2026},
      eprint={2410.22307},
      archivePrefix={arXiv},
      primaryClass={cs.LG},
      url={https://arxiv.org/abs/2410.22307}, 
}

@misc{cobbe2021trainingverifierssolvemath,
      title={Training Verifiers to Solve Math Word Problems}, 
      author={Karl Cobbe and Vineet Kosaraju and Mohammad Bavarian and Mark Chen and Heewoo Jun and Lukasz Kaiser and Matthias Plappert and Jerry Tworek and Jacob Hilton and Reiichiro Nakano and Christopher Hesse and John Schulman},
      year={2021},
      eprint={2110.14168},
      archivePrefix={arXiv},
      primaryClass={cs.LG},
      url={https://arxiv.org/abs/2110.14168}, 
}

@misc{chen2021evaluatinglargelanguagemodels,
      title={Evaluating Large Language Models Trained on Code}, 
      author={Mark Chen and Jerry Tworek and Heewoo Jun and Qiming Yuan and Henrique Ponde de Oliveira Pinto and Jared Kaplan and Harri Edwards and Yuri Burda and Nicholas Joseph and Greg Brockman and Alex Ray and Raul Puri and Gretchen Krueger and Michael Petrov and Heidy Khlaaf and Girish Sastry and Pamela Mishkin and Brooke Chan and Scott Gray and Nick Ryder and Mikhail Pavlov and Alethea Power and Lukasz Kaiser and Mohammad Bavarian and Clemens Winter and Philippe Tillet and Felipe Petroski Such and Dave Cummings and Matthias Plappert and Fotios Chantzis and Elizabeth Barnes and Ariel Herbert-Voss and William Hebgen Guss and Alex Nichol and Alex Paino and Nikolas Tezak and Jie Tang and Igor Babuschkin and Suchir Balaji and Shantanu Jain and William Saunders and Christopher Hesse and Andrew N. Carr and Jan Leike and Josh Achiam and Vedant Misra and Evan Morikawa and Alec Radford and Matthew Knight and Miles Brundage and Mira Murati and Katie Mayer and Peter Welinder and Bob McGrew and Dario Amodei and Sam McCandlish and Ilya Sutskever and Wojciech Zaremba},
      year={2021},
      eprint={2107.03374},
      archivePrefix={arXiv},
      primaryClass={cs.LG},
      url={https://arxiv.org/abs/2107.03374}, 
}

@misc{liu2023codegeneratedchatgptreally,
      title={Is Your Code Generated by ChatGPT Really Correct? Rigorous Evaluation of Large Language Models for Code Generation}, 
      author={Jiawei Liu and Chunqiu Steven Xia and Yuyao Wang and Lingming Zhang},
      year={2023},
      eprint={2305.01210},
      archivePrefix={arXiv},
      primaryClass={cs.SE},
      url={https://arxiv.org/abs/2305.01210}, 
}

@misc{zhou2023instructionfollowingevaluationlargelanguage,
      title={Instruction-Following Evaluation for Large Language Models}, 
      author={Jeffrey Zhou and Tianjian Lu and Swaroop Mishra and Siddhartha Brahma and Sujoy Basu and Yi Luan and Denny Zhou and Le Hou},
      year={2023},
      eprint={2311.07911},
      archivePrefix={arXiv},
      primaryClass={cs.CL},
      url={https://arxiv.org/abs/2311.07911}, 
}

@misc{qwen38,
    title = {{Qwen3.8-Max}: A New Bar for Coding and Cowork},
    url = {https://qwen.ai/blog?id=qwen3.8},
    author = {{Qwen Team}},
    month = {August},
    year = {2026}
}

@misc{museglimmerteam2026museglimmer30b,
      title={Muse Glimmer 30B: An On-Device Agentic Model}, 
      author={Muse Glimmer team},
      year={2026},
      eprint={2610.24628},
      archivePrefix={arXiv},
      primaryClass={cs.CL},
      url={https://arxiv.org/abs/2610.24628}, 
}

@misc{gemmateam2026gemma4,
  title         = {{Gemma 4 Technical Report}},
  author        = {{Gemma Team}},
  year          = {2026},
  eprint        = {2607.02770},
  archivePrefix = {arXiv},
  primaryClass  = {cs.CL},
  url           = {https://arxiv.org/abs/2607.02770}
}

@misc{vaswani2023attentionneed,
      title={Attention Is All You Need}, 
      author={Ashish Vaswani and Noam Shazeer and Niki Parmar and Jakob Uszkoreit and Llion Jones and Aidan N. Gomez and Lukasz Kaiser and Illia Polosukhin},
      year={2023},
      eprint={1706.03762},
      archivePrefix={arXiv},
      primaryClass={cs.CL},
      url={https://arxiv.org/abs/1706.03762}, 
}

@misc{lee2025programmingrefusalconditionalactivation,
      title={Programming Refusal with Conditional Activation Steering}, 
      author={Bruce W. Lee and Inkit Padhi and Karthikeyan Natesan Ramamurthy and Erik Miehling and Pierre Dognin and Manish Nagireddy and Amit Dhurandhar},
      year={2025},
      eprint={2409.05907},
      archivePrefix={arXiv},
      primaryClass={cs.LG},
      url={https://arxiv.org/abs/2409.05907}, 
}

@misc{fort2023scalinglawsadversarialattacks,
      title={Scaling Laws for Adversarial Attacks on Language Model Activations}, 
      author={Stanislav Fort},
      year={2023},
      eprint={2312.02780},
      archivePrefix={arXiv},
      primaryClass={cs.LG},
      url={https://arxiv.org/abs/2312.02780}, 
}

@inproceedings{10.1145/3696410.3714640,
author = {Wang, Tianlong and Jiao, Xianfeng and Zhu, Yinghao and Chen, Zhongzhi and He, Yifan and Chu, Xu and Gao, Junyi and Wang, Yasha and Ma, Liantao},
title = {Adaptive Activation Steering: A Tuning-Free LLM Truthfulness Improvement Method for Diverse Hallucinations Categories},
year = {2025},
isbn = {9798400712746},
publisher = {Association for Computing Machinery},
address = {New York, NY, USA},
url = {https://doi.org/10.1145/3696410.3714640},
doi = {10.1145/3696410.3714640},
booktitle = {Proceedings of the ACM on Web Conference 2025},
pages = {2562–2578},
numpages = {17},
location = {Sydney NSW, Australia},
series = {WWW '25}
}

@InProceedings{pmlr-v202-sheng23a,
  title = 	 {{F}lex{G}en: High-Throughput Generative Inference of Large Language Models with a Single {GPU}},
  author =       {Sheng, Ying and Zheng, Lianmin and Yuan, Binhang and Li, Zhuohan and Ryabinin, Max and Chen, Beidi and Liang, Percy and Re, Christopher and Stoica, Ion and Zhang, Ce},
  booktitle = 	 {Proceedings of the 40th International Conference on Machine Learning},
  pages = 	 {31094--31116},
  year = 	 {2023},
  volume = 	 {202},
  series = 	 {Proceedings of Machine Learning Research},
  month = 	 {23--29 Jul},
  publisher =    {PMLR},
}

@misc{gond2026llm42enablingdeterminismllm,
      title={LLM-42: Enabling Determinism in LLM Inference with Verified Speculation}, 
      author={Raja Gond and Aditya K Kamath and Ramachandran Ramjee and Ashish Panwar},
      year={2026},
      eprint={2601.17768},
      archivePrefix={arXiv},
      primaryClass={cs.LG},
      url={https://arxiv.org/abs/2601.17768}, 
}

@misc{zhao2023lingualinkeddistributedlargelanguage,
      title={LinguaLinked: A Distributed Large Language Model Inference System for Mobile Devices}, 
      author={Junchen Zhao and Yurun Song and Simeng Liu and Ian G. Harris and Sangeetha Abdu Jyothi},
      year={2023},
      eprint={2312.00388},
      archivePrefix={arXiv},
      primaryClass={cs.LG},
      url={https://arxiv.org/abs/2312.00388}, 
}

@misc{wang2026verillmlightweightframeworkpublicly,
      title={VeriLLM: A Lightweight Framework for Publicly Verifiable Decentralized Inference}, 
      author={Ke Wang and Zishuo Zhao and Xinyuan Song and Zelin Li and Libin Xia and Chris Tong and Bill Shi and Wenjie Qu and Eric Yang and Lynn Ai},
      year={2026},
      eprint={2509.24257},
      archivePrefix={arXiv},
      primaryClass={cs.CR},
      url={https://arxiv.org/abs/2509.24257}, 
}

@misc{song2024powerinferfastlargelanguage,
      title={PowerInfer: Fast Large Language Model Serving with a Consumer-grade GPU}, 
      author={Yixin Song and Zeyu Mi and Haotong Xie and Haibo Chen},
      year={2024},
      eprint={2312.12456},
      archivePrefix={arXiv},
      primaryClass={cs.LG},
      url={https://arxiv.org/abs/2312.12456}, 
}

@misc{zhang2025hardwaresoftwareplatforminference,
      title={Hardware and Software Platform Inference}, 
      author={Cheng Zhang and Hanna Foerster and Robert D. Mullins and Yiren Zhao and Ilia Shumailov},
      year={2025},
      eprint={2411.05197},
      archivePrefix={arXiv},
      primaryClass={cs.LG},
      url={https://arxiv.org/abs/2411.05197}, 
}

@misc{wimbauer2026fingerprintinginferencesystemslarge,
      title={Fingerprinting Inference Systems of Large Language Models}, 
      author={Anna Wimbauer and Jonas Möller and Erik Imgrund and Konrad Rieck},
      year={2026},
      eprint={2605.29979},
      archivePrefix={arXiv},
      primaryClass={cs.CR},
      url={https://arxiv.org/abs/2605.29979}, 
}

\appendix
\section{Optimization Algorithm}
\label{app:optimizer}

Algorithm~\ref{alg:eos-optimizer} details the constrained
optimizer used for both $J_1$ and $J_2$.
We distinguish the native faithful activation
$h_{\mathrm{ref}}$, used for final budget enforcement,
from the prefix activation $h_{\mathrm{obj}}$,
used for candidate realization and objective evaluation.
For $J_1$, both equal $h_i^f$.
For $J_2$, $h_{\mathrm{ref}}=h_i^f$ and
$h_{\mathrm{obj}}=\tilde{h}_i^P$, with the surrogate
header activations supplied as fixed context.

Here, $\operatorname{ActivationGradient}
(h_{\mathrm{obj}},\delta,J)$ denotes the gradient
obtained by backpropagating the EOS objective through
the BF16 cast to the FP32 pre-cast prefix activation
$\operatorname{FP32}(h_{\mathrm{obj}})+\delta$,
without differentiating through the realization
operator $\mathcal{R}$.

The realization operator $\mathcal{R}(h,u,\rho)$ returns
an FP32 additive perturbation from a unit-norm direction
$u$. It checks feasibility by adding the perturbation
to the FP32-converted activation $h$, casting the result
to BF16, and measuring the resulting relative activation
difference. If necessary, it reduces the perturbation
magnitude and retries.
During search, candidates are realized using
$h_{\mathrm{obj}}$.
The optimizer retains the highest-scoring direction
across all restarts and iterations.
Before returning, it realizes this direction again
using $h_{\mathrm{ref}}$ to obtain the final perturbation
$\delta^\star$.
The budget in Equation~\eqref{eq:attack-budget-constraint}
is additionally checked on the actual transmitted
activation during evaluation.

\begin{algorithm}[t]
\caption{Constrained EOS Optimizer}
\label{alg:eos-optimizer}
\begin{algorithmic}[1]
\Require Reference activation $h_{\mathrm{ref}}$,
         objective prefix $h_{\mathrm{obj}}$,
         budget $\rho$, objective $J$,
         restart count $R$, iterations $M$,
         backtracking limit $L$, tolerance $\epsilon$
\Ensure Feasible perturbation $\delta^\star$

\State $s^\star \gets -\infty$
\For{$r=1,\ldots,R$}
    \State Sample a random unit direction $u$
    \State $u \gets
        \arg\max_{w\in\{u,-u\}}
        J(\mathcal{R}(h_{\mathrm{obj}},w,\rho))$
    \State $\delta \gets
        \mathcal{R}(h_{\mathrm{obj}},u,\rho)$
    \State $s \gets J(\delta)$
    \If{$s>s^\star$}
        \State $(u^\star,s^\star) \gets (u,s)$
    \EndIf

    \For{$m=1,\ldots,M$}
        \State $\delta \gets
            \mathcal{R}(h_{\mathrm{obj}},u,\rho)$
        \State $s \gets J(\delta)$
        \State $g \gets
            \Call{ActivationGradient}
            {h_{\mathrm{obj}},\delta,J}$
            \Comment{Gradient at the FP32 pre-cast prefix}
        \State $v \gets g-\langle g,u\rangle u$

        \If{$\|v\|_F \leq \epsilon$
            \textbf{or}
            $\neg\operatorname{isfinite}(\|v\|_F)$}
            \State \textbf{continue}
        \EndIf
        \State $v \gets v/\|v\|_F$

        \For{$\ell=0,\ldots,L-1$}
            \State $\eta \gets 2^{-\ell}$
            \State $u' \gets
                (u+\eta v)/\|u+\eta v\|_F$
            \State $\delta' \gets
                \mathcal{R}(h_{\mathrm{obj}},u',\rho)$
            \State $s' \gets J(\delta')$

            \If{$s'>s$}
                \State $u \gets u'$
                \If{$s'>s^\star$}
                    \State $(u^\star,s^\star)
                        \gets (u',s')$
                \EndIf
                \State \textbf{break}
            \EndIf
        \EndFor
    \EndFor
\EndFor
\State $\delta^\star \gets
    \mathcal{R}(h_{\mathrm{ref}},u^\star,\rho)$
\State \Return $\delta^\star$
\end{algorithmic}
\end{algorithm}

\section{Complete Verification Outcomes}
\label{app:verification-outcomes}
\begin{table*}[t]
\centering
\caption{Complete verification outcomes across three benchmarks.
All values are percentages.
AS is inapplicable to clean executions; clean AF denotes
false alarms.
Arrows indicate the attacker's preferred direction
for perturbed executions.}
\label{tab:verification-outcomes-full}
\small
\setlength{\tabcolsep}{5.1pt}
\begin{tabular}{ll*{12}{r}}
\toprule
& & \multicolumn{3}{c}{TopLoc}
  & \multicolumn{3}{c}{SENTINEL}
  & \multicolumn{3}{c}{Activation-DiFR}
  & \multicolumn{3}{c}{TAO} \\
\cmidrule(lr){3-5}
\cmidrule(lr){6-8}
\cmidrule(lr){9-11}
\cmidrule(lr){12-14}
Model & Perturbation
& Accept $\uparrow$ & \multicolumn{2}{c}{Reject}
& Accept $\uparrow$ & \multicolumn{2}{c}{Reject}
& Accept $\uparrow$ & \multicolumn{2}{c}{Reject}
& Accept $\uparrow$ & \multicolumn{2}{c}{Reject} \\
\cmidrule(lr){4-5}
\cmidrule(lr){7-8}
\cmidrule(lr){10-11}
\cmidrule(lr){13-14}
& & & AS $\downarrow$ & AF $\uparrow$
    & & AS $\downarrow$ & AF $\uparrow$
    & & AS $\downarrow$ & AF $\uparrow$
    & & AS $\downarrow$ & AF $\uparrow$ \\
\midrule

Qwen3.8-27B & Clean
& 100.00 & --- & 0.00
& 12.22 & --- & 87.78
& 100.00 & --- & 0.00
& 100.00 & --- & 0.00 \\
& Gaussian
& 100.00 & 0.00 & 0.00
& 11.94 & 0.00 & 88.06
& 100.00 & 0.00 & 0.00
& 100.00 & 0.00 & 0.00 \\
& Uniform
& 100.00 & 0.00 & 0.00
& 12.22 & 0.00 & 87.78
& 100.00 & 0.00 & 0.00
& 100.00 & 0.00 & 0.00 \\
& \textbf{Ours}
& 52.22 & \textbf{0.00} & 47.78
& 4.44 & \textbf{0.00} & 95.56
& 1.11 & \textbf{0.00} & 98.89
& 4.44 & \textbf{0.00} & 95.56 \\
\midrule

Muse-Glimmer-30B & Clean
& 100.00 & --- & 0.00
& 97.78 & --- & 2.22
& 100.00 & --- & 0.00
& 100.00 & --- & 0.00 \\
& Gaussian
& 100.00 & 0.00 & 0.00
& 97.78 & 0.00 & 2.22
& 99.72 & 0.00 & 0.28
& 99.72 & 0.00 & 0.28 \\
& Uniform
& 100.00 & 0.00 & 0.00
& 97.78 & 0.00 & 2.22
& 100.00 & 0.00 & 0.00
& 100.00 & 0.00 & 0.00 \\
& \textbf{Ours}
& 100.00 & \textbf{0.00} & 0.00
& 97.78 & \textbf{0.00} & 2.22
& 1.11 & \textbf{0.00} & 98.89
& 52.22 & \textbf{0.00} & 47.78 \\
\midrule

Gemma-4-31B-IT & Clean
& 100.00 & --- & 0.00
& 28.89 & --- & 71.11
& 100.00 & --- & 0.00
& 100.00 & --- & 0.00 \\
& Gaussian
& 100.00 & 0.00 & 0.00
& 30.56 & 5.56 & 63.89
& 1.67 & 32.50 & 65.83
& 53.06 & 46.94 & 0.00 \\
& Uniform
& 100.00 & 0.00 & 0.00
& 28.61 & 5.56 & 65.83
& 1.11 & 36.67 & 62.22
& 48.89 & 51.11 & 0.00 \\
& \textbf{Ours}
& 0.00 & \textbf{0.00} & 100.00
& 32.22 & 5.56 & 62.22
& 0.00 & \textbf{0.00} & 100.00
& 0.00 & 100.00 & 0.00 \\

\bottomrule
\end{tabular}
\end{table*}
Table~\ref{tab:verification-outcomes-full} reports
verification outcomes for clean execution, random
perturbation baselines, and \textsc{Hush-Up}, using
identical watchdog configurations across conditions.
For Qwen, four of the 360 records in each condition
exceed SENTINEL's supported reference length.
We count these records as accepted rather than
successfully verified.

\myparagraph{Gemma Error Concentration and TAO Outcomes}
To characterize TAO's model-dependent attribution
outcomes, we analyze benign A100--H100 discrepancies
across hidden dimensions using 50 inputs.
The top $1\%$ of hidden dimensions account for a
median of $86.6\%$ of the squared discrepancy in
Gemma, compared with $36.6\%$ in Qwen and $14.7\%$
in Muse.
The two largest-error dimensions alone account for
$79.2\%$ in Gemma, compared with $31.9\%$ in Qwen
and $2.0\%$ in Muse.

This concentration highlights a distinction between
the global relative $\ell_2$ budget and TAO's
distributional checks.
The budget constrains aggregate perturbation magnitude
without requiring the error distribution to remain
within TAO's absolute- and relative-error percentile
envelopes.
All optimized Gemma perturbations satisfy the budget
but trigger rejection at the true injection boundary,
yielding $100\%$ attribution success in the main
experiments.
Same-budget random perturbations are also rejected
in a substantial fraction of cases despite their
limited effect on task accuracy, consistent with
sensitivity to error distribution beyond aggregate
magnitude.

\section{Complete Injection-Depth Results}
\label{app:injection-depth}

Table~\ref{tab:injection-depth-full} reports task accuracy
and attribution success for TAO and TopLoc across all
evaluated injection depths. TopLoc achieves zero
attribution success for every model and depth.

\begin{table*}[t]
\centering
\caption{Complete results for the effect of injection depth
on GSM8K task accuracy and attribution success (AS).
Each setting uses 30 problems with one decoding trial
per problem and a separately calibrated perturbation
budget for the corresponding injection boundary.
All values are percentages.}
\label{tab:injection-depth-full}
\small
\setlength{\tabcolsep}{10.1pt}
\begin{tabular}{lrrrrrrrrr}
\toprule
& \multicolumn{3}{c}{Qwen3.8-27B}
& \multicolumn{3}{c}{Muse-Glimmer-30B}
& \multicolumn{3}{c}{Gemma-4-31B-IT} \\
\cmidrule(lr){2-4}
\cmidrule(lr){5-7}
\cmidrule(lr){8-10}
Depth
& Acc.\ $\downarrow$ & TAO AS $\downarrow$ & TopLoc AS $\downarrow$
& Acc.\ $\downarrow$ & TAO AS $\downarrow$ & TopLoc AS $\downarrow$
& Acc.\ $\downarrow$ & TAO AS $\downarrow$ & TopLoc AS $\downarrow$ \\
\midrule
20\%
& 0.00 & 0.00 & 0.00
& 3.33 & 0.00 & 0.00
& 6.67 & 100.00 & 0.00 \\
40\%
& 0.00 & 0.00 & 0.00
& 0.00 & 0.00 & 0.00
& 0.00 & 46.67 & 0.00 \\
60\%
& 0.00 & 0.00 & 0.00
& 0.00 & 0.00 & 0.00
& 0.00 & 30.00 & 0.00 \\
80\%
& 0.00 & 0.00 & 0.00
& 0.00 & 0.00 & 0.00
& 0.00 & 80.00 & 0.00 \\
100\%
& 0.00 & 0.00 & 0.00
& 0.00 & 0.00 & 0.00
& 0.00 & 100.00 & 0.00 \\
\bottomrule
\end{tabular}
\end{table*}

\section{Additional Experimental Details}
\label{app:experimental-details}

\myparagraph{Inference Templates}
We use each model's native inference template.
We set the reasoning effort to low for Qwen and Muse
and enable reasoning mode for Gemma.
Qwen's opening \texttt{<think>} marker is included
in the prefill, whereas Muse and Gemma generate their
reasoning headers autoregressively.

\myparagraph{Budget Calibration}
We use 200 fixed inputs from the GSM8K training set
to profile benign cross-hardware numerical variation,
with 160 used for calibration and 40 held out for validation.

\myparagraph{Decoding Configuration}
We use a sampling temperature of $0.7$ and top-$p$
sampling with $p=0.95$ for autoregressive generation.
The maximum generation length is 1,024 tokens for
GSM8K and HumanEval+, and 2,048 tokens for IFEval.

\myparagraph{Optimization Configuration}
For \textsc{Hush-Up}, we use $\tau=0.99$.
For the EOS optimizer, we use $R=4$ random restarts,
$M=100$ optimization iterations per restart, and at most
$L=6$ backtracking steps per iteration.
The optimization objective is computed at temperature $1$.
We construct the perturbation $\delta$ in FP32 and
enforce the budget on the activation difference realized
after BF16 rounding. We set the tangent-gradient norm tolerance to
$\epsilon=10^{-30}$.

\myparagraph{Watchdog Implementation}

For TopLoc, Activation-DiFR, and TAO, we first check
the final boundary and, upon rejection, apply the discrepancy test at the known injection boundary. For SENTINEL, we preserve its upstream-first order:
we check the injection boundary first, followed by the final boundary if the first check accepts. 
We use TopLoc's official proof-generation and verification API. For Activation-DiFR, we train detectors using the official framework and prescribed procedure. We independently implement TAO's discrepancy tests as specified in the original work. For SENTINEL, we retain its four distance metrics and IQR-based threshold calibration, adapting its training-oriented mechanism to variable-length inference using a position-wise EMA. The EMA references and thresholds are frozen during evaluation. 

\section{Separate Evaluation of the Surrogate $J_2$ Objective}
\label{app:j2-analysis}

We separately evaluate $J_2$ at the applicable boundaries of the injection-depth experiment (Section~\ref{sec:results}), using the same problems, decoding configuration, perturbation budgets, and optimization hyperparameters.
We evaluate Muse and Gemma, excluding Qwen because
its opening header is included in the prefill rather
than generated autoregressively. The synthetic prompt preserves the public chat template
and a fixed generic system message, with its user message
filled with repetitions of the token for ``the'' so that
the total prompt length matches the observed prefill length.
Table~\ref{tab:j2-depth} reports the results on GSM8K.

For Muse, $J_2$ achieves $100\%$ Stop$\le$30
and zero task accuracy at the $40\%$, $60\%$,
and $80\%$ boundaries.
For Gemma, it achieves the same results at the
$40\%$ and $60\%$ boundaries, while Stop$\le$30
decreases to $63.33\%$ and accuracy rises to $16.67\%$
at the $80\%$ boundary.
The $16.67\%$ accuracy at Gemma's $80\%$ boundary
does not imply uniformly intact responses.
Some correct responses contain
garbled text or malformed channel markers, or continue repetitively until the generation
limit despite containing the correct answer.
Thus, answer accuracy alone does not capture
the degradation in response quality.
These results show that the surrogate objective
can induce premature termination at deeper worker
boundaries, although its effectiveness varies
with the model and injection depth.
We omit the $100\%$ boundary because no downstream
stage remains there, making $J_2$
independent of the perturbation (Section~\ref{sec:optimization}).

\begin{table}[t]
\centering
\caption{Separate evaluation of $J_2$ on GSM8K.
All values are percentages.
Depth denotes the injection boundary.
All settings match the injection-depth experiment.
The $100\%$ boundary is inapplicable to $J_2$.}
\label{tab:j2-depth}
\small
\setlength{\tabcolsep}{4pt}
\begin{tabular*}{\columnwidth}{
    @{\extracolsep{\fill}}l*{4}{r}@{}
}
\toprule
& \multicolumn{2}{c}{Muse-Glimmer-30B}
& \multicolumn{2}{c}{Gemma-4-31B} \\
\cmidrule(lr){2-3}
\cmidrule(lr){4-5}
Depth
& Stop$\le$30 $\uparrow$ & Acc.\ $\downarrow$
& Stop$\le$30 $\uparrow$ & Acc.\ $\downarrow$ \\
\midrule
Clean
& 0.0 & 93.3
& 0.0 & 93.3 \\
40\%
& 100.0 & 0.0
& 100.0 & 0.0 \\
60\%
& 100.0 & 0.0
& 100.0 & 0.0 \\
80\%
& 100.0 & 0.0
& 63.3 & 16.7 \\
\bottomrule
\end{tabular*}
\end{table}

\section{Qualitative Generation Examples}
\label{app:qualitative-examples}

Table~\ref{tab:qualitative-examples} presents eight
representative generation traces under \textsc{Hush-Up}.
The first four are drawn from the main experiments,
while the remaining four come from the separate
$J_2$ evaluation at $80\%$ depth.
The examples illustrate diverse behaviors, including
premature termination, persistent abnormal generation,
and recovery with or without successful task completion.

\begin{table}[!htbp]
\centering
\caption{Representative generation traces under
\textsc{Hush-Up}. Non-English spans are replaced
with language labels, and ellipses indicate omissions.
\texttt{[EOS]} denotes a model-specific terminating token.
Token counts include generated headers and terminating
tokens, when present.}
\label{tab:qualitative-examples}
\small
\setlength{\tabcolsep}{3pt}
\begin{tabular}{
    @{}p{0.10\columnwidth}
    p{0.24\columnwidth}
    p{0.59\columnwidth}@{}
}
\toprule
Model & Outcome & Representative generation \\
\midrule
Qwen
& Immediate termination
& \texttt{[EOS]}
  \newline
  \textit{1 token; no answer.} \\
\midrule
Muse
& Post-header termination
& \texttt{to=self<|message|>[EOS]}
  \newline
  \textit{4 tokens; no answer.} \\
\midrule
Gemma
& Short-output termination
& \texttt{<|channel>thought}
  \newline
  \texttt{*~~~Topic:[EOS]}
  \newline
  \textit{8 tokens; fails strict prompt-level evaluation.} \\
\midrule
Gemma
& Repetitive generation
& \texttt{// own own own own own \ldots}
  \newline
  \textit{1,024 tokens; length limit reached,
  no valid answer.} \\
\midrule
Gemma
& Abnormal generation followed by recovery
& \texttt{io}[Chinese] [Japanese] [Chinese]
  \newline
  [Portuguese] [Korean] \texttt{la la la} [Thai]
  \newline
  \ldots\ [Mixed]\ \ldots
  \newline
  Wait, I see the [Mixed] of a glitch.
  Let me restart.
  \newline
  \ldots
  \newline
  Final Answer: 5
  \newline
  \textit{198 tokens; correct answer.} \\
\midrule
Gemma
& Persistent abnormal generation
& \texttt{C\&A:}
  \newline
  \ldots
  \newline
  [Arabic] \ldots\ [Korean]
  \newline
  \ldots
  \newline
  \texttt{**the** **the** **the** \ldots}
  \newline
  [Thai] [Thai] [Thai] \ldots
  \newline
  \textit{1,024 tokens; length limit reached,
  no valid answer.} \\
  \midrule
Gemma
& Attempted recovery followed by termination
& [Mixed]
  \newline
  \ldots
  \newline
  Actually, let'[Korean]'s calculate:
  \newline
  \texttt{**[EOS]}
  \newline
  \textit{55 tokens; no answer.} \\
\midrule
Gemma
& Recovery followed by repetitive generation
& [Mixed]
  \newline
  \ldots
  \newline
  Let me start over. I need to solve the problem
  provided by the [Chinese] user.
  \newline
  \ldots
  \newline
  Final answer: 45
  \newline
  \ldots
  \newline
  \texttt{Final answer: 45<channel|>}
  \newline
  To find the total amount of money in the
  [Burmese] [Burmese] [Burmese] \ldots
  \newline
  \textit{1,024 tokens; correct answer,
  but generation reaches the length limit.} \\
\bottomrule
\end{tabular}
\end{table}

\end{document}